\documentclass[a4paper,fleqn,usenatbib]{mnras}

\usepackage[T1]{fontenc}
\usepackage{ae,aecompl}
\usepackage{tablefootnote}

\usepackage{graphicx}	
\usepackage{amsmath}	
\usepackage{amssymb}	
\usepackage{url}

\title[Blazhko effect in the Galactic bulge]{Blazhko effect in the Galactic bulge fundamental mode RR~Lyrae stars I: Incidence rate and differences between modulated and non-modulated stars}

\author[Prudil\,\&\,Skarka]{
Z. Prudil$^{1,2}$\thanks{prudilz@ari.uni-heidelberg.de}, M. Skarka$^{3}$\thanks{marek.skarka@csfk.mta.hu}
\\
$^{1}$ Department of Theoretical Physics and Astrophysics, Masaryk University, Kotl\'{a}\v{r}sk\'{a} 2, 611 37 Brno, Czech Republic\\
$^{2}$ Astronomisches Rechen-Institut, Zentrum f{\"u}r Astronomie der Universit{\"a}t at Heidelberg, M{\"o}nchhofstr. 12-14, \\ ~~69120 Heidelberg, Germany\\
$^{3}$ Konkoly Observatory, Research Centre for Astronomy and Earth Sciences, Hungarian Academy of Sciences, \\ ~~Konkoly Thege Mikl\'{o}s \'{u}t 15-17, H-1121 Budapest,Hungary \\
}

\date{Accepted XXX. Received YYY; in original form ZZZ}

\pubyear{2016}

\begin{document}
\label{firstpage}
\pagerange{\pageref{firstpage}--\pageref{lastpage}}
\maketitle

\begin{abstract}
We present the first paper of a series focused on the Blazhko effect in RR Lyrae type stars pulsating in the fundamental mode, that are located in the Galactic bulge. A~comprehensive overview about the incidence rate and light-curve characteristics of the Blazhko stars is given. We analysed 8\,282 stars having the best quality data in the OGLE-IV survey, and found that at least $40.3$\,\% of stars show modulation of their light curves. The number of Blazhko stars we identified is 3\,341, which is the largest sample ever studied implying the most relevant statistical results currently available. Using combined data sets with OGLE-III observations, we found that 50\,\% of stars that show unresolved close peaks to the main component in OGLE-IV are actually Blazhko stars with extremely long periods. Blazhko stars with modulation occur preferentially among RR Lyrae stars with shorter pulsation periods in the Galactic bulge. Fourier amplitude and phase coefficients based on the mean light curves appear to be substantially lower for Blazhko stars than for stars with unmodulated light curve in average. We derived new relations for the compatibility parameter $D_{m}$ in $I$ passband and relations that allow for differentiating modulated and non-modulated stars easily on the basis of $R_{31}$, $\phi_{21}$ and $\phi_{31}$. Photometric metallicities, intrinsic colours and absolute magnitudes computed using empirical relations are the same for Blazhko and non-modulated stars in the Galactic bulge suggesting no correlation between the occurrence of the Blazhko effect and these parameters. 

\end{abstract}


\section{Introduction}\label{Sect:Introduction}

Horizontal-branch stars, that cross the classical instability strip, are known as RR Lyrae (RRL) stars. These radially pulsating stars have applications in many astrophysical fields \citep[e.g. distance indicators,][]{catelan2008} and constitute the most numerous class of catalogued pulsating stars in our Galaxy and close galactic systems \citep[e.g.][]{soszynski2014,soszynski2016}. Usually they are sorted according to their light curve shapes, which result from the mode in which they pulsate, into three basic historical groups: RRab (asymmetric light curve, fundamental mode), RRc (more symmetric light curve, first-overtone mode), RRd (fuzzy light curve, simultaneous fundamental and first-overtone mode pulsations). 

Substantial portion of the RRab stars shows from days to thousands-of-days long amplitude and phase modulation of their light curves known as the Blazhko (BL) effect \citep{blazhko1907}. The mechanism standing behind the BL effect is still a subject to discussion. Recent discoveries of the dynamical phenomena \citep[such as period doubling, e.g.][]{szabo2010} suggest that resonances between low- and high-order radial modes could play important role in explanation of the BL effect \citep{kollath2011,buchler2011}. In addition, it seems that only modulated RRab stars show additional radial and non-radial pulsational modes \citep{benko2014,szabo2014}. 

However, we are still at the beginning of understanding the BL effect, and no connection between physical characteristics of RRL stars and modulation has been reported so far. Simply it is not known why some of the RRL stars show modulation and other with similar parameters not. In addition, there is some indication that BL effect could actually be a temporary, episodic phase in RRL life \citep[e.g.][]{sodor2007,jurcsik2012}.

The estimates of the incidence rate of BL stars depend strongly on the data quality and time span of the data, number of investigated stars, and also differ for different stellar systems \citep[see table 1 in][]{kovacs2016}. For Galactic field RRLs, recent estimates are between 35 and 60\,\% \citep[based on precise space data,][]{benko2014,szabo2014}. This discrepancy shows that these estimates could be somewhat misleading due to small number of stars available for the statistics.

There are many other unanswered questions surrounding the Blazhko phenomenon. \citet{alcock2003}, \citet{jurcsik2011} and \citet{skarka2014b} reported on slightly shorter pulsation periods of BL stars in comparison with non-BL stars, while \citet{moskalik2003} found no difference. Some studies showed that modulated stars are slightly less luminous \citep{jurcsik2011,skarka2014b}, but some show the opposite \citep[e.g.][]{arellano2012}. Contradictory results are also reported for metallicity estimates: \citet{moskalik2003} suggested that the occurrence rate of BL stars increases with increasing metallicity, while \citet{smolec2005} and \citet{skarka2014b} found no metallicity dependence. However, the mentioned studies are based on various number of stars with different number of observations with different quality, in addition, in diverse stellar systems, which could intrinsically differ. Such comparison is, therefore, somewhat questionable. It is clear that for reliable statistics of any kind a large and homogeneous sample of stars with good quality data is desired to reduce the selection effects and reasonably describe overall characteristics of BL stars in particular stellar system.

We utilized high-quality measurements of more than 8\,000 RRL stars from the Galactic bulge (GB) provided by the fourth generation of the Optical Gravitational Lensing Experiment \citep[OGLE-IV,][]{udalski2015}, and searched for the modulation. This data set is almost perfect to secure reliable statistical results on the BL effect: the observations are homogeneous (gathered with the same telescope and through a single filter), have very good quality, sufficient time span (four years + 12 years from OGLE-II and OGLE-III if needed in some targets), and all observed stars belong to one, essentially homogeneous group of stars. After the identification of the BL stars, we compare their periods and light curve characteristics with those of non-BL stars. 

At this point it is good to note that we only identified BL stars and did not perform detail analysis of the modulation properties within this study. Modulation periods, amplitudes and relations between them, will be elaborated in the second paper of the series by \citet[][in prep.]{skarka2017}. 

The article is organized as follows. Section \ref{Sect:SampleSelection} describes the data and sample selection criteria used in this investigation. Methods for analysis of the observations are detailed in sect. \ref{Sect:Analysis}. The results on occurrence rate and differences between BL and non-modulated stars are presented in sect.~\ref{Sect:Results}. The results and future prospects are summarized in sect. \ref{Sect:Summary}.

\section{Sample selection}\label{Sect:SampleSelection}  

The OGLE survey is an ongoing long-term experiment managed by University of Warsaw, Poland\footnote{\url{ http://ogle.astrouw.edu.pl/}}. Since its beginning in early nineties it underwent a few substantial upgrades defining different phases of the survey \citep{udalski1992,udalski1997,udalski2003,udalski2015}. The present phase of the project, OGLE-IV, started in 2010 at Las Campanas in Chile. An 1.3-meter telescope equipped with a 262.5 megapixel 32-chip mosaic CCD camera ({\emph V} and {\emph I} photometric filters) is recently used for observation of the LMC, SMC, GB, and Galactic disc \citep{udalski2015}.

\begin{figure*}
	\includegraphics[width=\columnwidth]{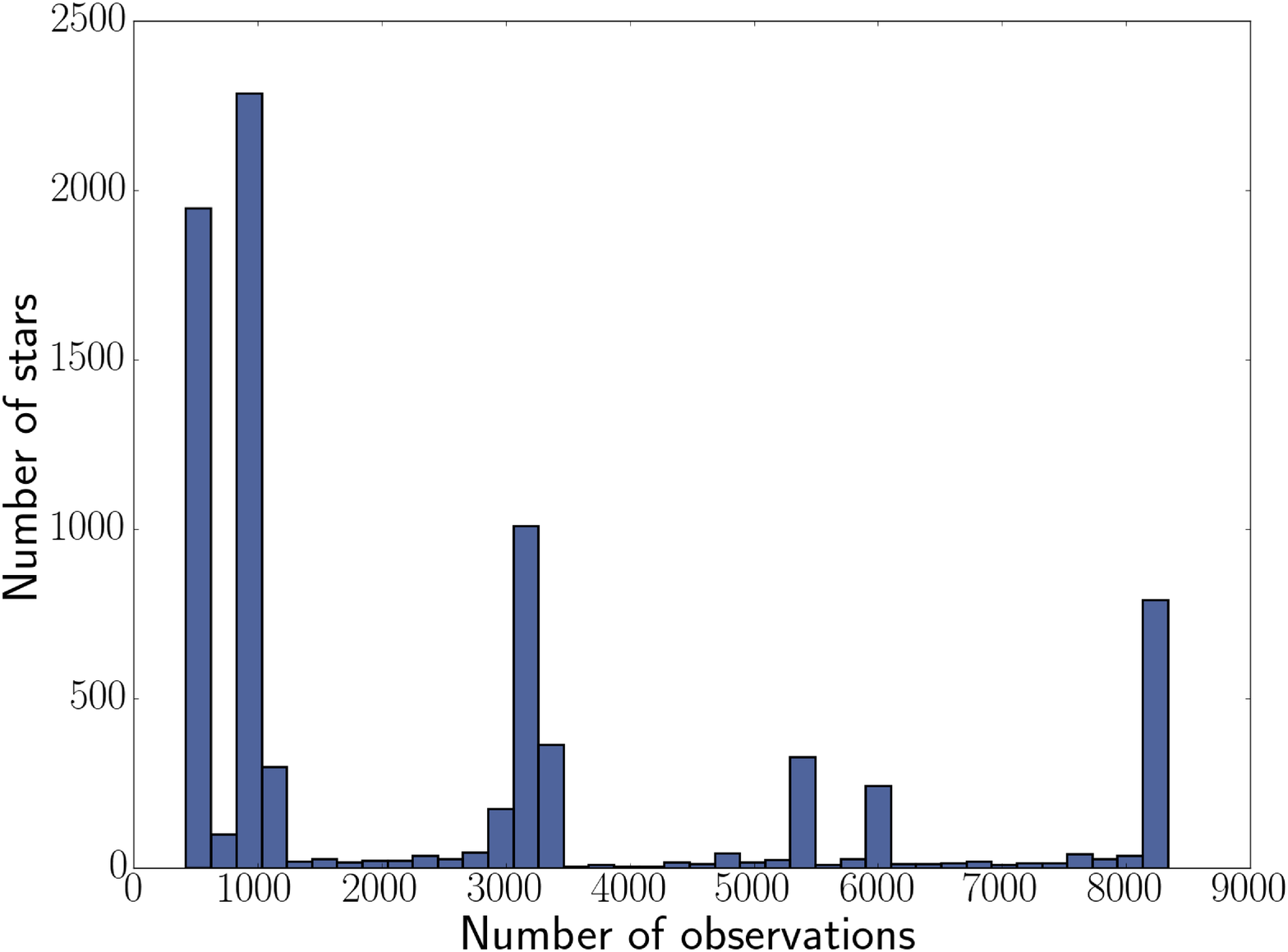}\includegraphics[width=\columnwidth]{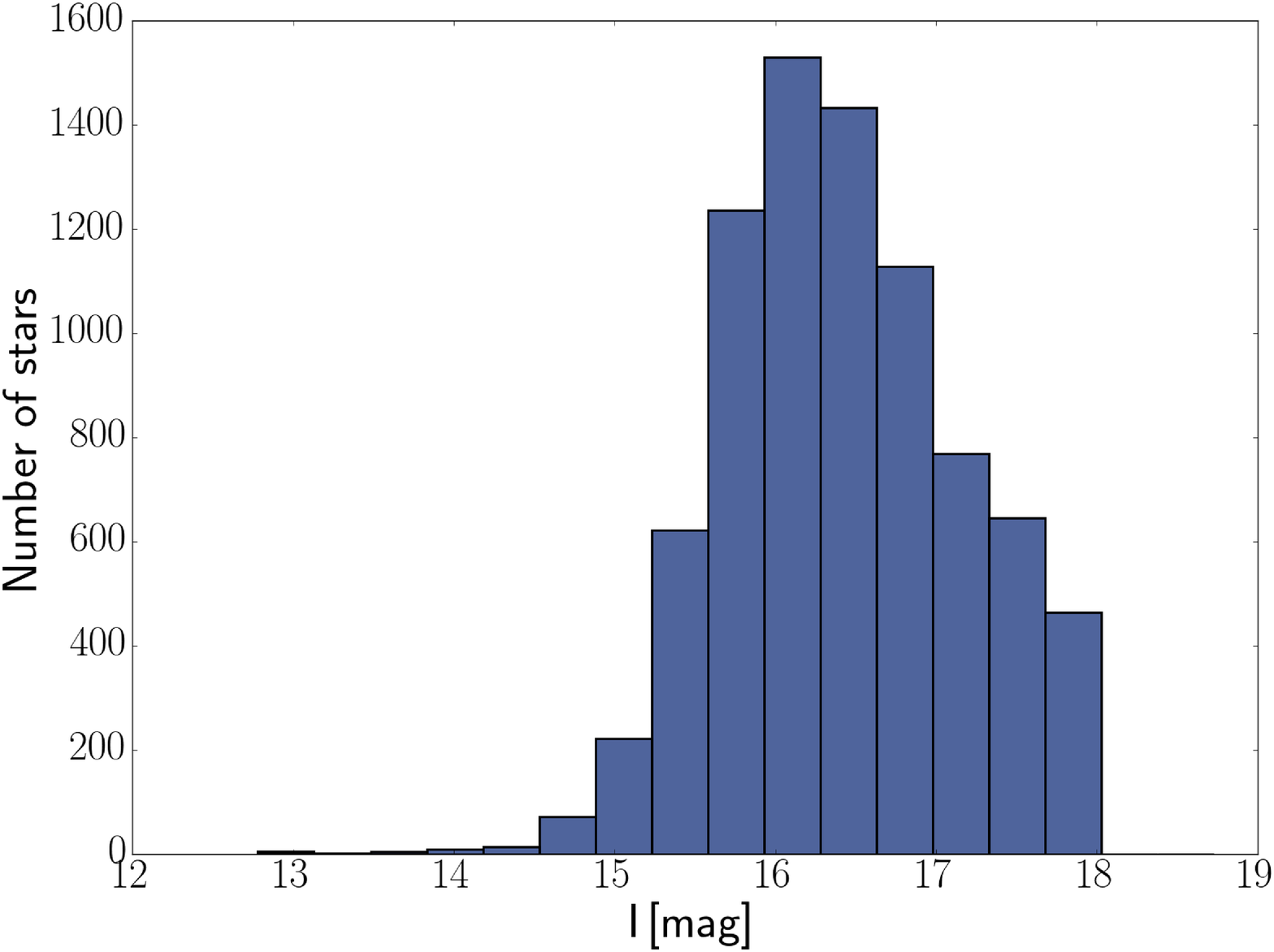}
    \caption{Distribution of stars according to the number of observations (the left-hand panel) and the mean magnitude (the right-hand panel).}
    \label{Fig:NumberOfPointsDistribution}
\end{figure*} 

In 2014 the first data for more than 38\,000 RRL type stars (27\,258 of RRab type) located in the GB gathered in OGLE-IV were publicly released. Together with photometric data, \citet{soszynski2014} published a complete catalogue with light ephemerides, mean brightness, peak-to-peak amplitudes and Fourier coefficients that we adopted for further analysis\footnote{\url{ftp://ftp.astrouw.edu.pl/ogle/ogle4/OCVS/blg/rrlyr/}}. 

Not all of available observations were suitable for our purposes. Because we aimed to search for the BL effect on sufficiently large sample of stars, we had to take into account the quality of the observations and number of points for particular star, secure sufficient number of stars to get reliable statistics, use only stars located in the GB. The following criteria were set up as the most convenient to comply all our requirements:
\begin{enumerate}
	\item We used only measurements in {\it I}-band because they contain substantially more data points than the observations in {\it V} pass band, and have much better quality. 
	\item Stars with Galactic latitude below $b=-8^{\circ}$ were ignored (possible membership of the Sagittarius tidal stream).
	\item Stars from globular clusters were omitted.
	\item Stars identified by \citet{hajdu2015} as binary candidates were removed from the sample.
	\item We did not use RRL stars listed in {\it remarks.txt} file (stored at the FTP mirror), which contains RRL with uncertain  identification.
	\item After visual inspection we decided to use only stars brighter than 18\,mag in {\it I} because of acceptable scatter.
	\item Stars with suspicious light curves that were probably misclassified as RRab were ignored.  
	\item Only stars with more than 420 points (size of the data file of 10\,kB) were used.
\end{enumerate} 

Our final sample contains 8\,282 stars that have between 420 and 8\,334 data points (median 1\,016). The left panel of Fig.~\ref{Fig:NumberOfPointsDistribution} shows the distribution of observations per star. The `discrete' pattern of the histogram is because some fields were observed more often than the others. We can see that the majority of stars has less than about 1\,200 data points. The right-hand panel of the same figure shows the distribution of main magnitudes of the sample stars. Because we did not remove any of the bright stars from the sample, it is likely that very few of the brightest stars are actually foreground Galactic field RRLs.

The mean photometric error of a single point for stars with mean $I<15.5$\,mag is 4\,mmag, for faint stars with $I\sim 18$\,mag it is about 20 mmag. The time span ({\it T$_{S}$}) of the observations is similar for all stars: from 1\,100 to 1\,400\,d (median 1334 days).

\section{Searching for the BL effect}\label{Sect:Analysis}

\begin{figure*}
	\includegraphics[width=0.5\columnwidth]{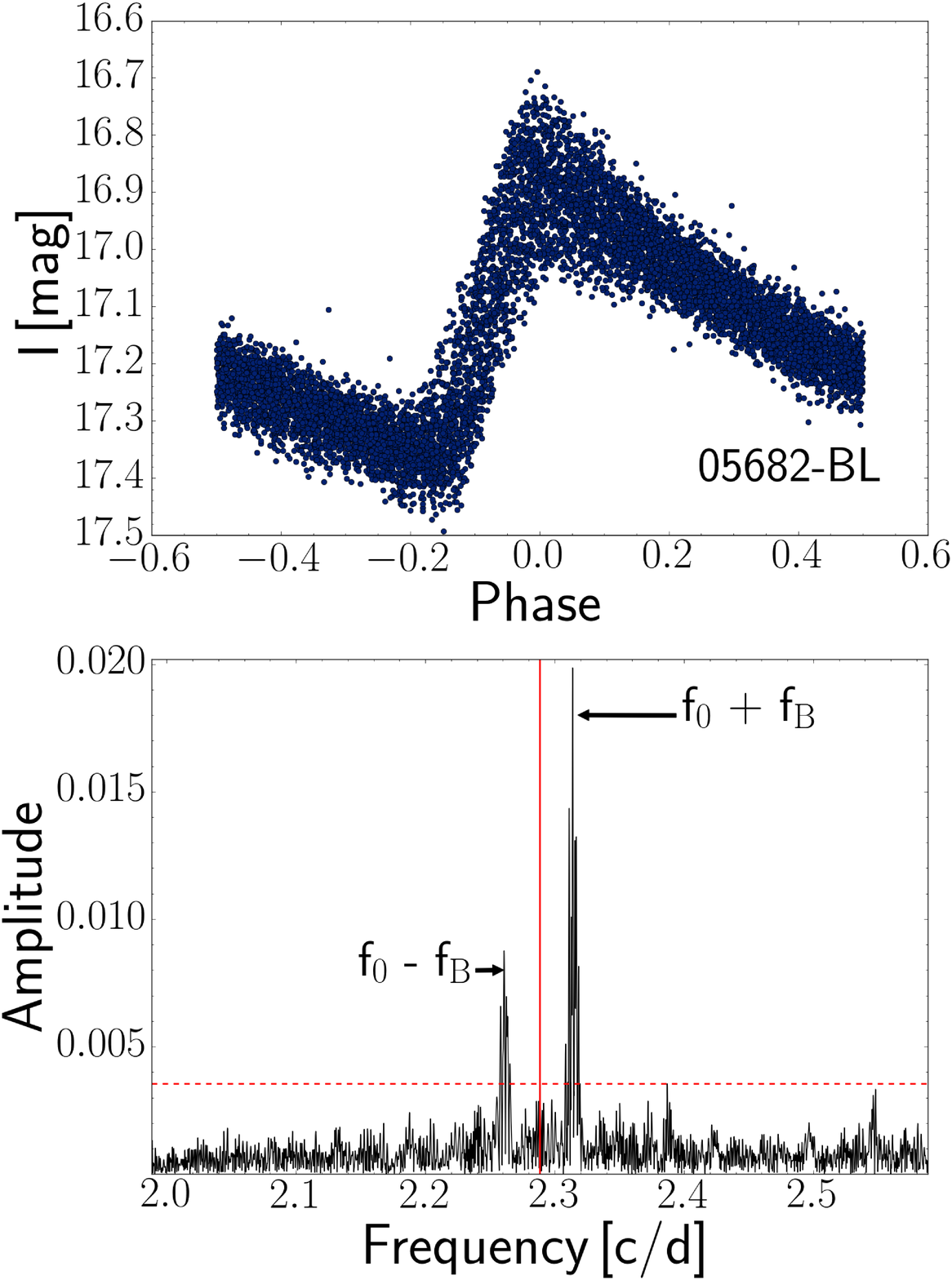}\includegraphics[width=0.5\columnwidth]{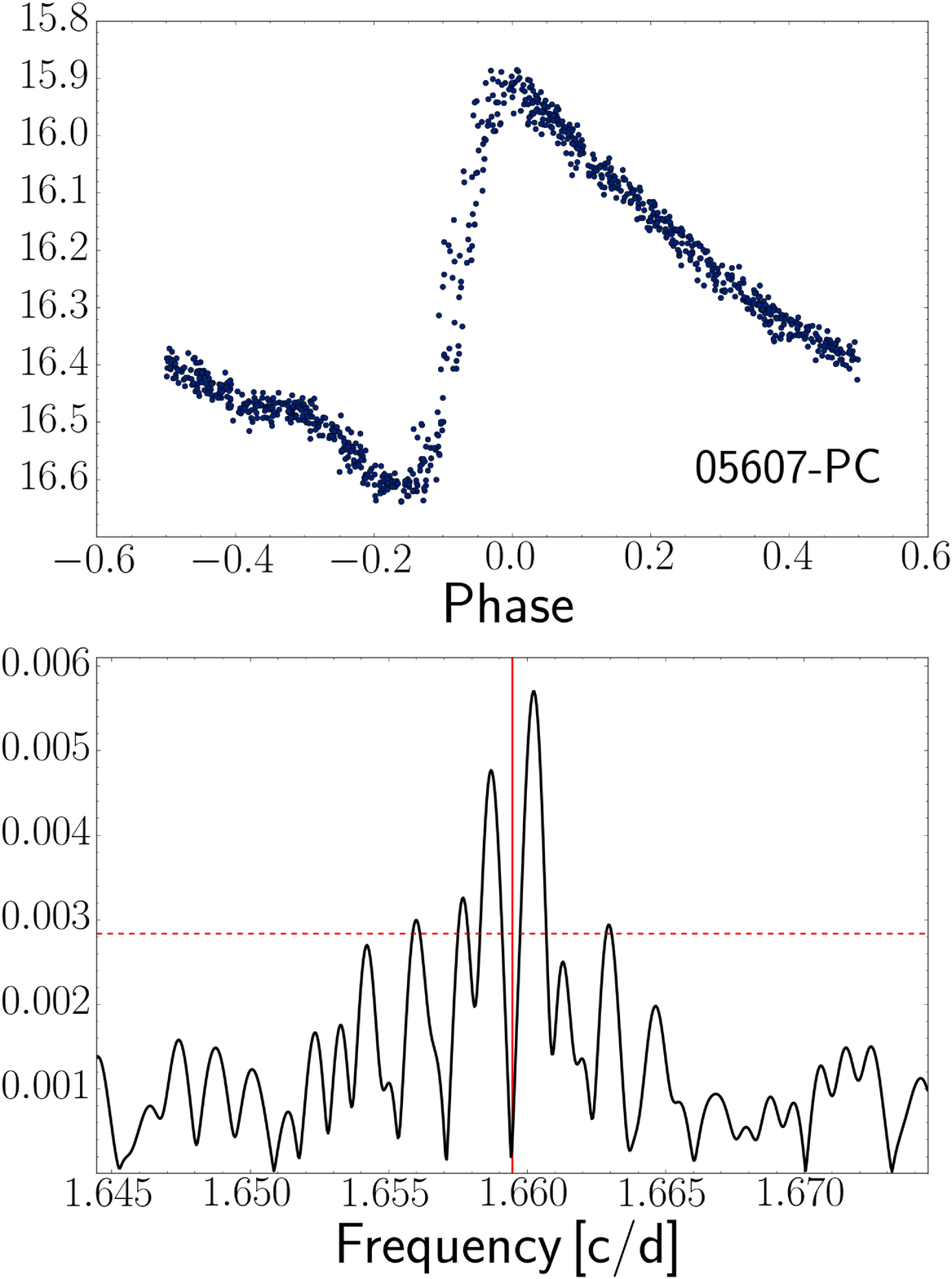}\includegraphics[width=0.5\columnwidth]{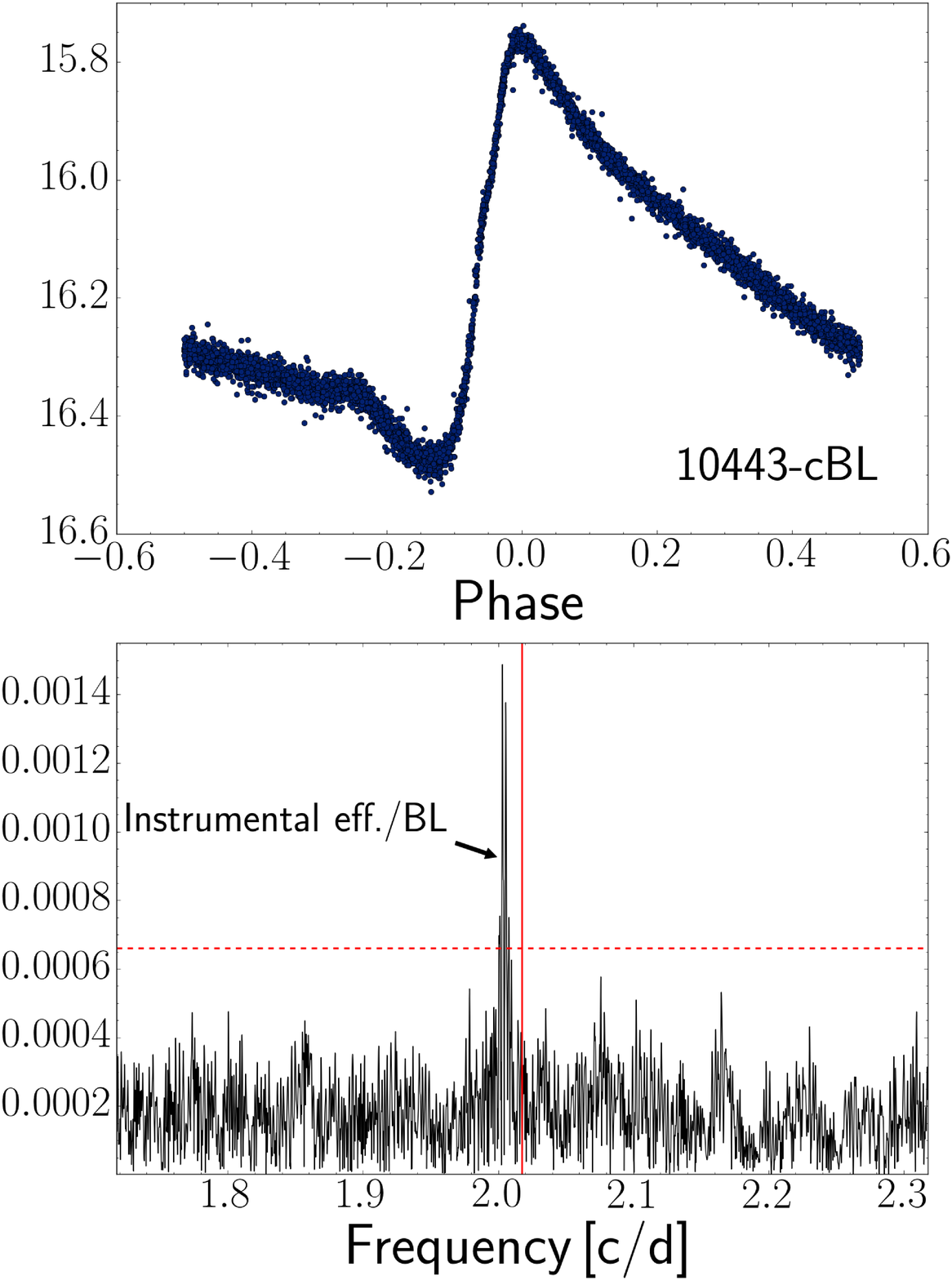}\includegraphics[width=0.5\columnwidth]{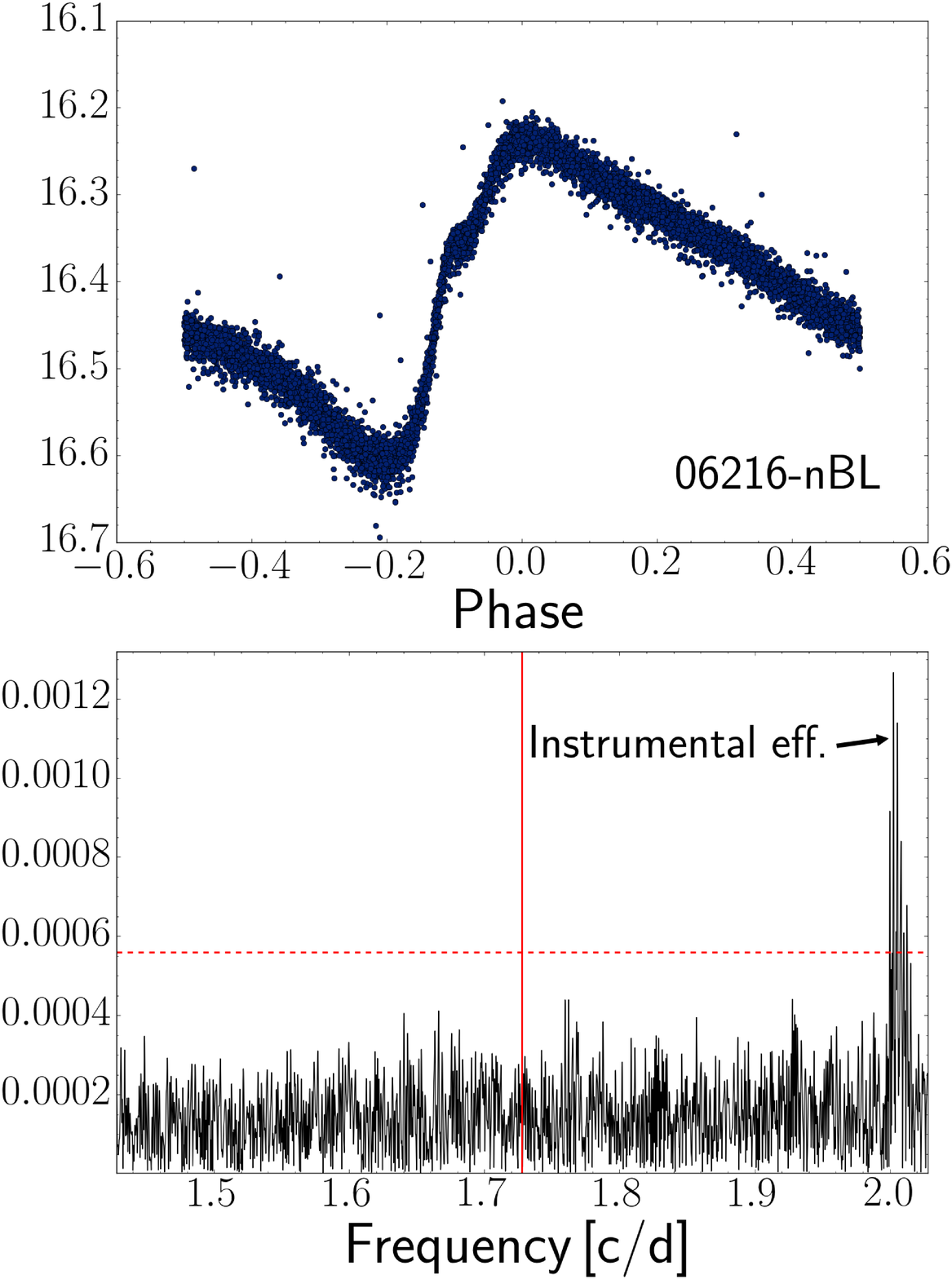}
    \caption{Examples of phase curves and frequency spectra around the main pulsation frequency after prewhitening with ten pulsation harmonics. From left to the right BL, PC, candidate, and unmodulated stars are shown. The horizontal dotted lines show the average {\it S/N}$=3.5$ limit and the red vertical lines mark possition of the main pulsation frequency.}
    \label{Fig:SortingSample}
\end{figure*}

First we plotted raw observations and phase curves \citep[using pulsation periods determined by][]{soszynski2014} and visually inspected all light curves. Although very subjective and time-consuming, this approach allowed for making a basic idea about the data and assigning each star to one of the three preliminary categories (clear BL effect, candidate, non-modulated star). It also helped to identify special cases \citep[for example, discovery of peculiar pulsation modes in several RRLs,][]{smolec2016,prudil2016} and stars showing distinct seasonal changes and secular trends in the mean magnitude suggesting blends. It should be borne in mind that this procedure was only informative having no influence on further analysis and decision about the presence of the BL effect. After the visual inspection, we decided not to remove any of the light-curve outliers because their number was generally low and their separation from the mean light curve was mostly negligible (see Fig.~\ref{Fig:SortingSample}).

Because the most prominent signs of modulation, which is detectable also in scattered data, are the side peaks close to the basic pulsation frequency ($f_{0}$) and its multiples \citep[$kf_{0}$, for details see][]{szeidl2009,szeidl2011}, we searched for the BL effect in the frequency spectra of the time series. The distance between side peak and the pulsation frequency defines the modulation frequency. 

The side peaks would be most easily and quickly identified using fully automatic procedures employing classical Fourier prewhitening techniques. Although this approach would be very comfortable and spare a lot of time, we must remember that  BL effect could be very complex, and that OGLE is a ground-based single-side survey producing data affected by various instrumental effects resulting in various aliases and false peaks in the frequency spectrum (typically the frequency at $\sim 2.005$\,c/d, which is frequency corresponding to twice the sidereal day (see Fig.~\ref{Fig:SortingSample}); frequencies at integer multiples of 1/day; higher noise in low-frequency range, etc.). Proper identification of the modulation is, therefore, not a straightforward task. The advantages and efficiency of individual analysis against fully automated procedures were well demonstrated by \citet{skarka2014a} and \citet{skarka2016b}. 

However, owing to the number of stars for analysis and time necessary for the individual analysis, it would be extremely time consuming to perform manual step by step prewhitening. Thus, we chose the way of the golden mean. We automatically prewhitened frequency spectrum of each of the sample stars with first ten pulsation {\bf harmonics} using non-linear least-squares methods, and then searched manually for the side peaks near $f_{0}$ and $2f_{0}$ in residuals using \textsc{Period04} software \citep{lenz2004}. Because our goal was only to identify the BL effect, the manual (non)identification of the side peaks was quite fast. We considered side peak as significant when it had signal-to-noise ratio ({\it S/N}) larger than 3.5\footnote{The {\it S/N} is simply defined as the ratio of the amplitude of the measured peak and the average of the amplitudes of peaks in $\pm 1$\,c/d vicinity of the measured peak.}.

We adopted similar criteria as introduced in \citet{skarka2016b} for assigning the stars to individual categories. Regarding the (non)detection of the side peaks and their characteristics our categories are as follows:
\begin{enumerate}
	\item {\it BL stars} -- stars with detectable side peak(s) within $f_{0}\pm 0.3$\,c/d, but no closer than 1.5/{\bf {\it T$_{S}$}} \citep{nagy2006,skarka2016b}. As a BL star we also considered stars that showed unambiguous secular amplitude variations with side peaks indicating modulation period longer than 2/3 {\it T$_{S}$}. An example of frequency spectrum of a BL star after prewhitening with ten pulsation harmonics in the vicinity of $f_{0}$ is shown in the most-left panel of Fig.~\ref{Fig:SortingSample}.
	\item {\it Stars with long-term/irregular period changes (PC)} -- detected side peak(s) were closer than 1.5/{\it T$_{S}$}, no apparent amplitude variations in the light curve (the middle-left panel of Fig.~\ref{Fig:SortingSample}).
	\item {\it Candidates (cBL)} -- stars that have close side peak(s) with $S/N<3.5$, or side peaks with $S/N>3.5$ near integer multiples of cycles per day within our limits for BL stars, or showing many peaks near the $S/N\sim3.5$ limit where we were unable to decide which peak could be the consequence of the modulation. Candidate stars are typically stars with periods close to 0.5 days where possible BL peaks interfere with instrumental peaks (the middle-right panel of Fig.~\ref{Fig:SortingSample}).
	\item {\it Non-modulated (nBL)} -- Signs of modulation undetectable (Fig.~\ref{Fig:SortingSample}, the right-hand panel).
\end{enumerate}

\begin{figure}
	\includegraphics[width=\columnwidth]{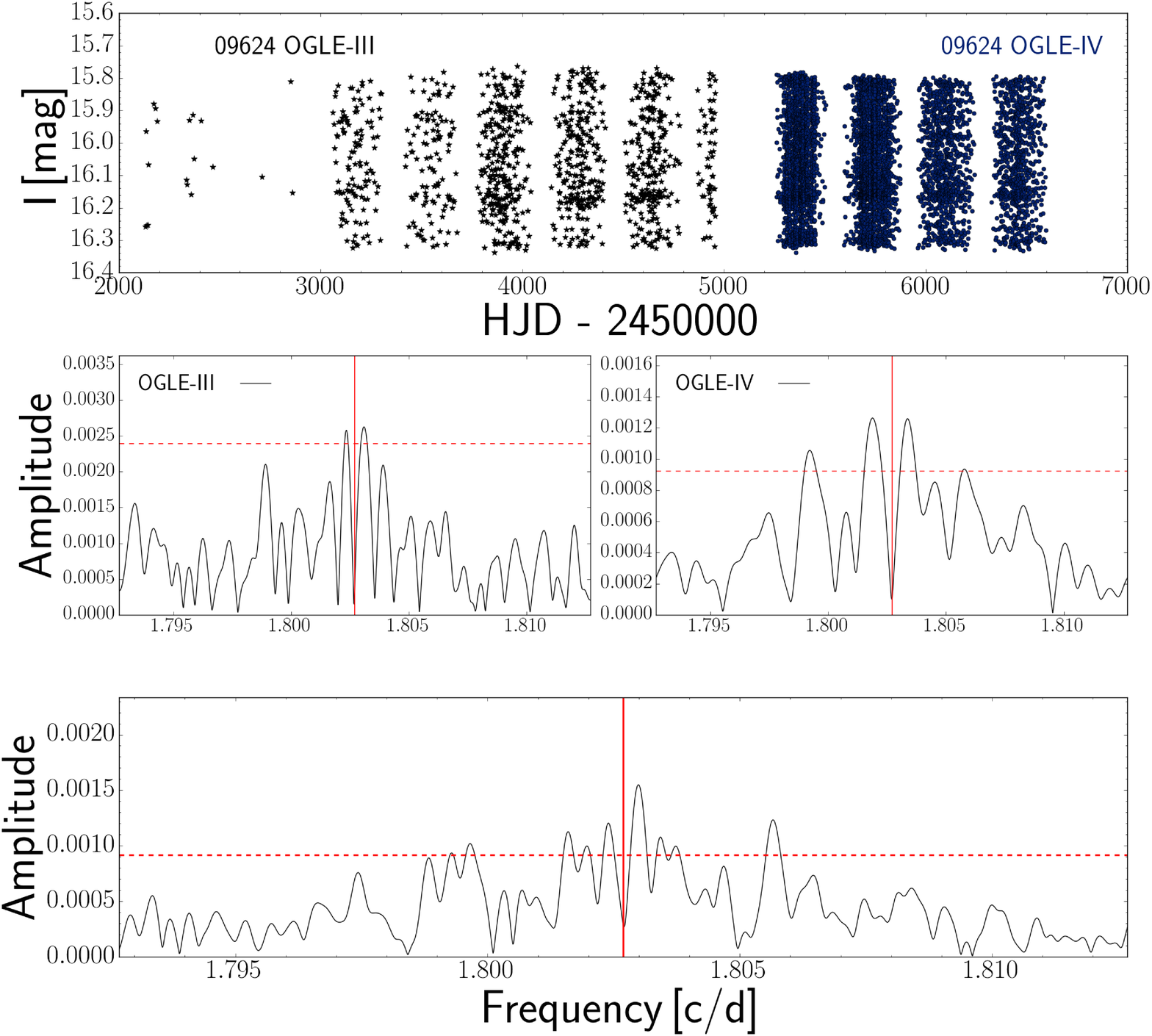}
    \caption{Combined OGLE-III and OGLE-IV dataset (top-panel) and corresponding frequency spectra (bottom panel). It is seen that based purely on the OGLE-IV observations (middle right-hand panel) the BL nature, represented by a tiny amplitude variation, could be very hardly revealed. Dashed red horizonta lines show the average {\it S/N}$=3.5$ limit. The vertical red lines mark the possition of the main pulsation freqeuncy.}
    \label{Fig:PC_Example}
\end{figure}

Stars that showed side peaks closer than 1.5/{\it T$_{S}$} were searched for additional data in OGLE-III survey to unambiguously decide about PC/BL membership. Because OGLE-III observations usually contain significantly less points than OGLE-IV, we combined both data sets, which is allowed by the close similarity of the data \citep{soszynski2014}. First we determined zero points for both datasets by applying 6-order Fourier fit. Then we shifted and combined the datasets, prewhitened 6 basic pulsation harmonics and analysed the residuals. With combined dataset we were able to determine BL periods safely up to the length of 3\,000 days, in some stars up to 4\,000 days. In a few cases the amplitudes were significantly different for OGLE-IV and OGLE-III data. These stars were not analysed and remained in the PC class. An example of this procedure is shown in Fig.~\ref{Fig:PC_Example}, which shows BL star with extremely long BL effect.

Finally, the light curves of all BL and candidate stars were visually inspected once again to reveal secular or seasonal changes suggesting possible blends, or some instrumental problems. Additional light from a close star would change the amplitude and the mean magnitude of the target star mimicking the BL effect (for example, due to variable seeing). Such stars would show distinct peak in the low-frequency range with higher amplitude than possible side peaks. We searched for such peaks in the range between 0 and 0.2\thinspace c/d in all BL and candidate stars and identified possible problems with blends in 21 stars (including visually identified and non-modulated stars, marked with `b' in the last column of Table \ref{Tab:ListOfStars}).

\section{Results}\label{Sect:Results}

We used only BL and nBL stars for comparison between BL and non-modulated stars, candidate and PC stars were omitted from the investigation. Thus all statistics and figures are based purely on BL and non-modulated stars.

\subsection{Incidence rate of the BL stars}\label{Subsect:IncidenceRate}

Identification of the sample stars regarding four adopted groups is in Table \ref{Tab:ListOfStars}. We identified BL effect in 3\,341 of the target stars. This is almost twice as more as all cataloged BL stars identified in all stellar systems together so far. In 29 objects (`a' in the column Comments in Table \ref{Tab:ListOfStars}) we detected doublet in a vicinity of $f_{\rm 0}$ (close to the other harmonic frequencies it was not detected). Besides these objects, all of the stars identified as modulated can be considered as firm BL stars because at least two side peaks with equidistant spacing from $kf_{0}$ were detected in their frequency spectra, or unambiguous secular amplitude change is apparent in 29 stars with extremely long modulation (marked with `lm' in the last column of the Table~\ref{Tab:ListOfStars}).

\begin{table*}
\centering
\caption{Showcase of the structure of the full online table, which lists the membership to one of our groups: non-modulated star (nBL), candidate star (cBL), BL star (BL) and PC star (PC), and observability in K2 campaigns 9 and 11 (0 -- not on silicon, 1 -- near silicon, 2 -- on silicon). The name of a star is in a form of OGLE-BLG-RRLYR-ID. The Remark 'a' in the Rem column means that only one side peak was detected, 'b' means possible blend, `lm' means that modulation period is longer than 2/3{\it T$_{S}$}, `h' is when the side peaks were identified in higher pulsation harmonics first, letters `c' and `m' means that stars are present in \citet{collinge2006} and \citet{moskalik2003}. Furthermore, we included rise time, together with calculated metalicity, $(V-I)_{0}$ and absolute magnitude in {\it I}-band. Additional information about pulsation periods, amplitudes and Fourier coefficients can be found in \citet{soszynski2014}. The full table is available as a supplementary material to this paper. Amplitudes of the sidepeaks and modulation periods will occur in \citet{skarka2017}, in prep.}
\label{Tab:ListOfStars}
\begin{tabular}{lcccccccc}
\hline
ID    & Group & K2-9 & K2-11 & Rem & {\it RT} & [Fe/H] & $(V-I)_{0}$\,[mag]  & $M_{I}$\,[mag]       \\ \hline
00162 & BL    & 0    & 2     &     & 0.176  & -1.09      & 0.467 & 0.183 \\
01983 & PC    & 0    & 0     &     & 0.178  & -0.86      & 0.419 & 0.297 \\
01994 & nBL   & 0    & 0     &     & 0.145  & -0.57      & 0.441 & 0.375 \\
08688 & cBL   & 2    & 0     & b   & 0.123 & -0.98      & 0.406 & 0.274 \\
  \dots    &   \dots    &  \dots    &    \dots   &  \dots   &     \dots      &     \dots       &     \dots     &  \dots    \\ \hline   
\end{tabular}
\end{table*}

Only 26 stars were assigned to the candidate group (`cBL' in Table \ref{Tab:ListOfStars}), 53 other stars were identified as PC stars (`PC' in Table \ref{Tab:ListOfStars}). In the remaining stars no signs of modulation were detected (`nBL' in the second column of Table \ref{Tab:ListOfStars}). From the analysis of the combined datasets (i.e., OGLE-III and IV) we find that 50\,\% of stars identified as PC in four-years long OGLE-IV data are actually BL stars with extremely long modulation periods. This finding could have large impact on incidence rates that are estimated only on the basis of short data sets. 

The percentage of modulated stars is 40.3\,\%; 40.0\,\% when we omit stars with only one detected {\bf side} peak, respectively. The true portion and completeness of BL identification is extraordinary difficult to estimate because the amplitudes of the side peaks depend on the strength of the modulation both in amplitude and frequency/phase, which can be very diverse. In addition, the detectability of the side peaks depends on the noise level, which is unique for each star and data set. Also the human factor and used methods could play role -- some of the BL stars could be missed or misclassified. 

\begin{figure}
	\includegraphics[width=\columnwidth]{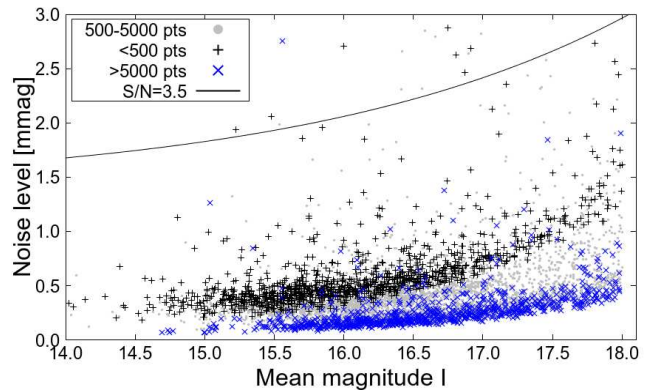}
    \caption{Noise level in Fourier amplitude spectra of non-blazhko stars from our sample as a function of the mean magnitude. Stars with less than 500 data points are shown with black crosses, stars with more than 5000 points are plotted with blue crosses to highlight the influence of number of points on the noise level. The solid line schematically shows $S/N=3.5$ limit for stars with less than 500 observations.}
    \label{Fig:IncidenceRateInvestigation}
\end{figure}

Figure \ref{Fig:IncidenceRateInvestigation} shows the noise level around the basic pulsation frequency for stars from our sample, that we identified as non-modulated, as a function of the mean magnitude. It is seen that the noise level increases with increasing mean magnitude, which could be naturally expected due to increasing scatter. The noise is also higher for stars with low number of points (black crosses). The solid line is an exponential fit of the noise level of stars with less than 500 observations multiplied by 3.5 that schematically shows the $S/N=3.5$ limit for the stars with the most noisy Fourier spectra. 

In the sample of BL stars observed by the {\it Kepler} space telescope, one fourth of the modulated stars has the side peaks with Fourier amplitudes lower than 1 mmag \citep{benko2014}. There are 1320 non-modulated stars that have $S/N=3.5$ limit larger than 0.001 mag in our sample. Possibly all of these stars could actually be BL stars in which the modulation was undetectable. If this were true, then these 1320 stars would constitute 28\,\% of all BL stars, and the full percentage of modulated stars in the GB would be 56\,\%. However, this is a huge simplification and a very strong assumption that stands only on a very limited sample of BL stars in the {\it Kepler} field.

The forty percent is high incidence rate in comparison with previous studies of GB. \citet{moskalik2003} estimated the percentage of modulated stars as 23\,\% (OGLE-I), \citet{mizerski2003} found 25\,\% of stars to be modulated (OGLE-II), \citet{collinge2006} reported 28 \,\% (OGLE-II), and \citet{soszynski2011} gives 30\,\% of RRab type stars exhibiting the BL effect (OGLE-II + OGLE-III). The rising tendency of the estimates with the longer, higher cadence and better quality datasets is obvious. The current estimates by \citet[][47\,\%]{jurcsik2009}; \citet[][43\,\%]{sodor2012}; \citet[][43\,\%]{nemec2013}; and \citet[][47\,\%]{szabo2014,benko2016} show that the occurrence rate is a bit higher for Galactic field stars than for GB stars. On one hand, the lower percentage of GB BL stars could be a natural property of this Galactic subsystem. Confirmation of this possibility is of high importance because it is still not known what physical processes rule the occurrence of the BL effect. On the other hand, the different result for the GB could be due to selection effects possibly caused by different quality data with different time span. Former studies used only low number of stars (maximally several tens) that could also distort the statistics. Unfortunately, we do not know which possibility is more likely, because no other study with similarly large and good-quality sample of stars has been available so far.

A more reliable picture of the efficiency of the detection of the BL effect and its incidence rate in GB will be available after the data gathered by mission {\it K2} \citep{howell2014} campaigns 9 and 11 will be analysed. More than two thousands of our sample stars fall on the silicon (1293 non-modulated stars, 910 BL stars, 13 PC stars and 9 candidates with flag `2', see the two last columns in the Table \ref{Tab:ListOfStars}). However, due to time-limited observations, this will tell us something only about BL stars with short modulation cycle.

\begin{figure*}
	\includegraphics[width=\columnwidth]{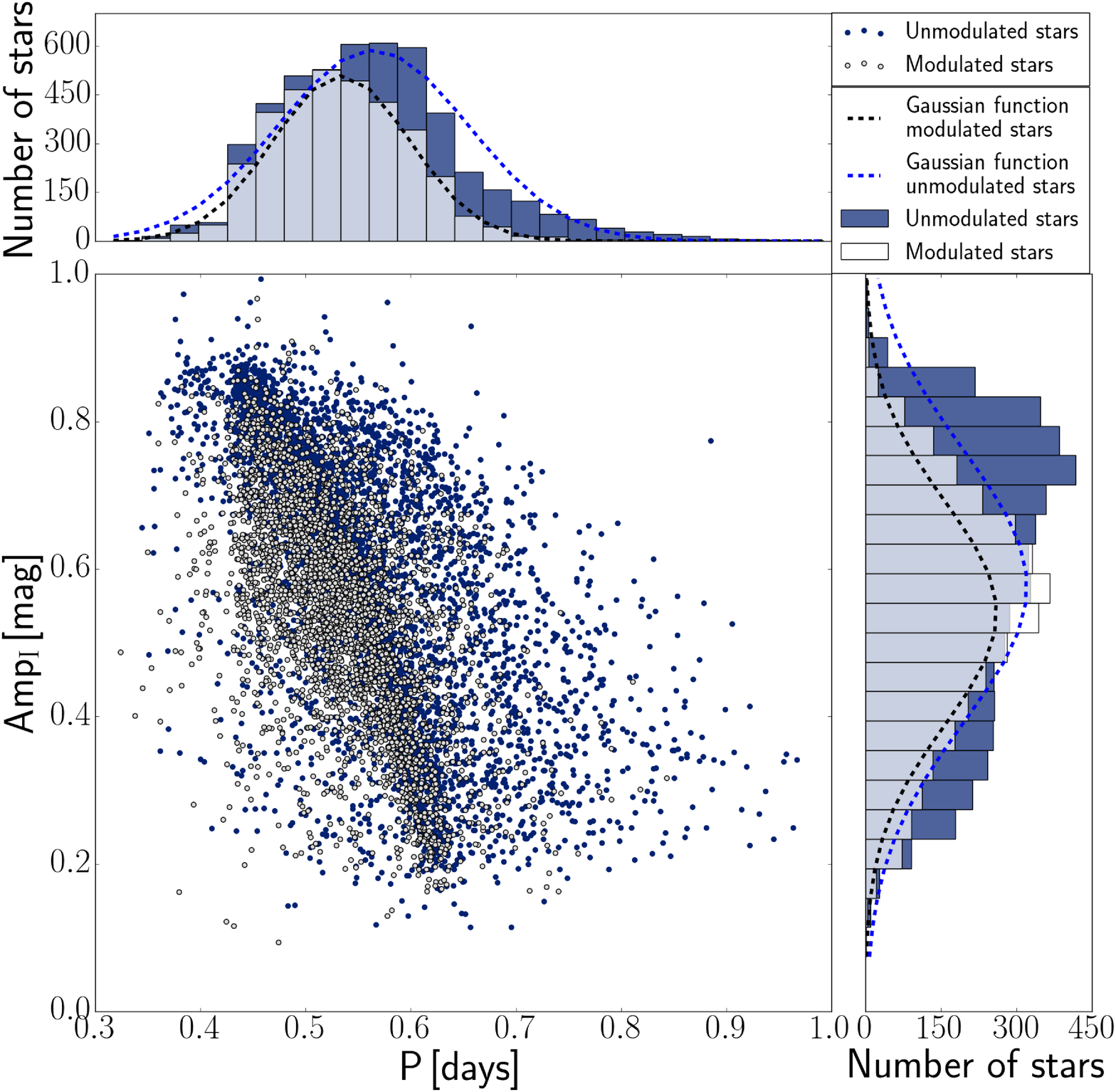}\includegraphics[width=\columnwidth]{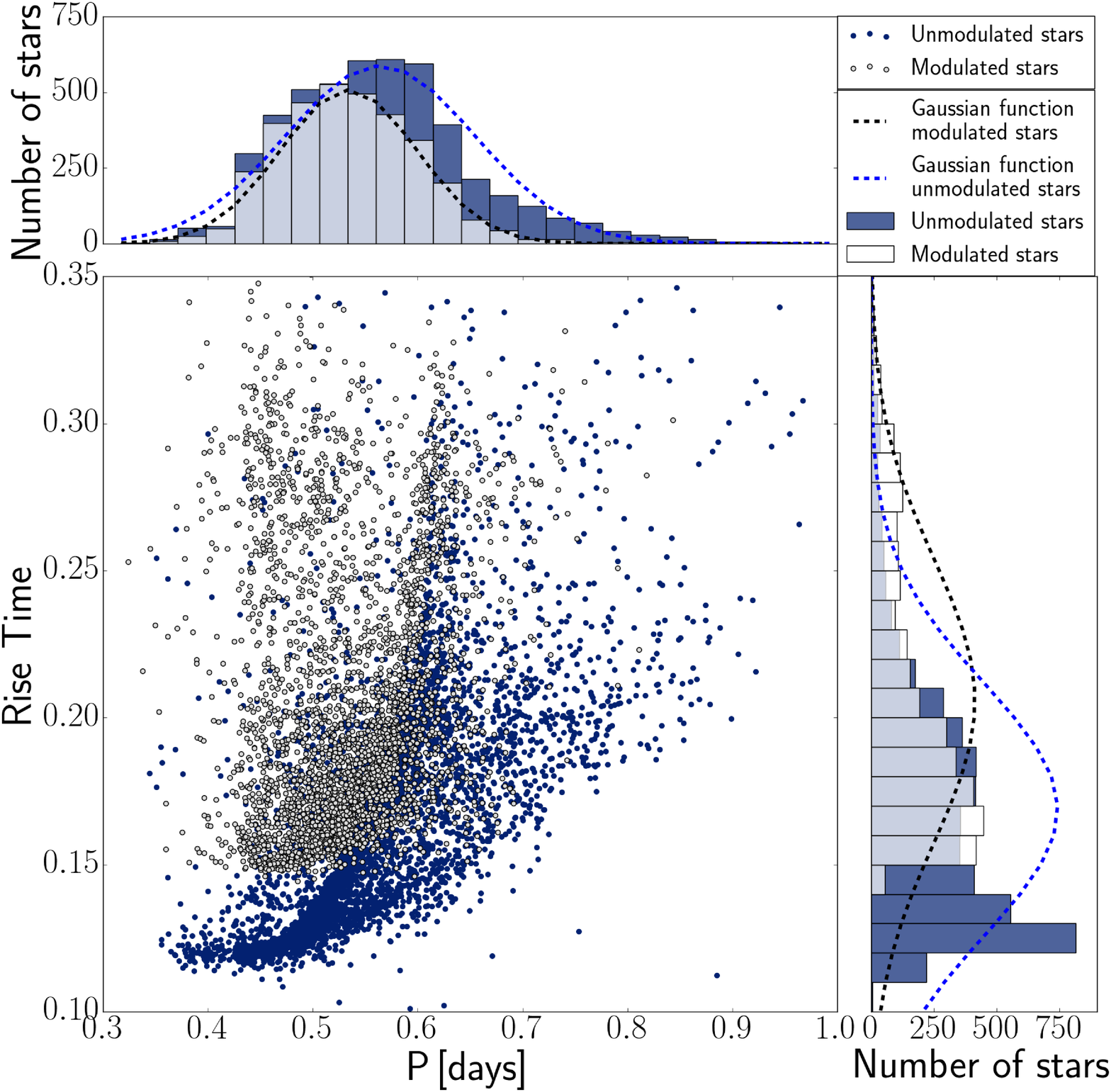}
    \caption{Period-amplitude (Bailey) diagram together with pulsation-period distribution and distribution of pulsational amplitudes (the left-hand panels) and plots showing the distribution of the rise time. Normal distributions with the same mean and standard deviation are plotted with lines in the histograms for comparison.}
    \label{Fig:Per-ampDistribution}
\end{figure*} 

\begin{table*}
	\centering
	\caption{Comparison of the mean parameters of single-period RRL stars and stars with the BL effect. The uncertainties in the last digits (in parentheses) are the standard 1 $\sigma$ errors of the means. Fourier parameters are based on cosine fit \citep{soszynski2014}. Metallicity, intrinsic colour and absolute magnitude show differences only for stars with $D_{m}<3$.}
	\label{Tab:Parameters}
	\begin{tabular}{lccccccccccc} 
		\hline
		 & $P_{\rm Puls}$\,[days] & $Amp_{I}$\,[mag] & $R_{21}$ & $R_{31}$ & $\phi_{21}$\,[rad] & $\phi_{31}$\,[rad] & $RT$ & [Fe/H] & $(V-I)_{0}$\,[mag] & $M_{I}$\,[mag]\\
		\hline
		nBL & 0.564(1) & 0.577(3)	 & 0.491(1) & 0.319(1) & 4.450(4) & 2.832(7) & 0.169(3) & -0.962(5) & 0.490(2) & 0.198(2) \\
		BL & 0.533(1) & 0.540(3) & 0.458(1) & 0.274(1) & 4.367(4) & 2.616(8) &	0.219(8) & -0.969(5) & 0.487(2) & 0.202(2) \\
		\hline
	\end{tabular}
\end{table*}

\begin{figure}
	\includegraphics[width=\columnwidth]{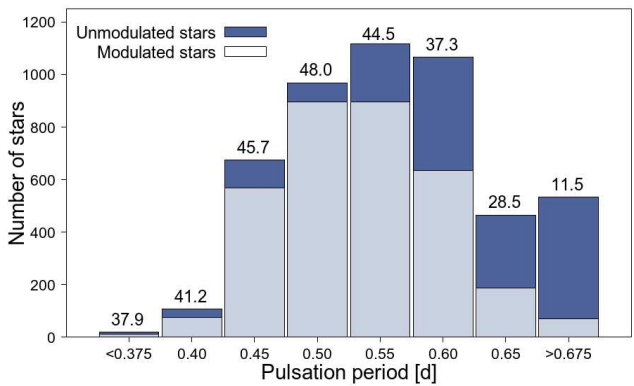}
    \caption{Distribution of the BL stars according to the pulsation period. Number above each bin shows the corresponding percentage of BL stars in given range.}
    \label{Fig:IncidenceRatePeriodBins}
\end{figure} 

\begin{figure}
	\includegraphics[width=\columnwidth]{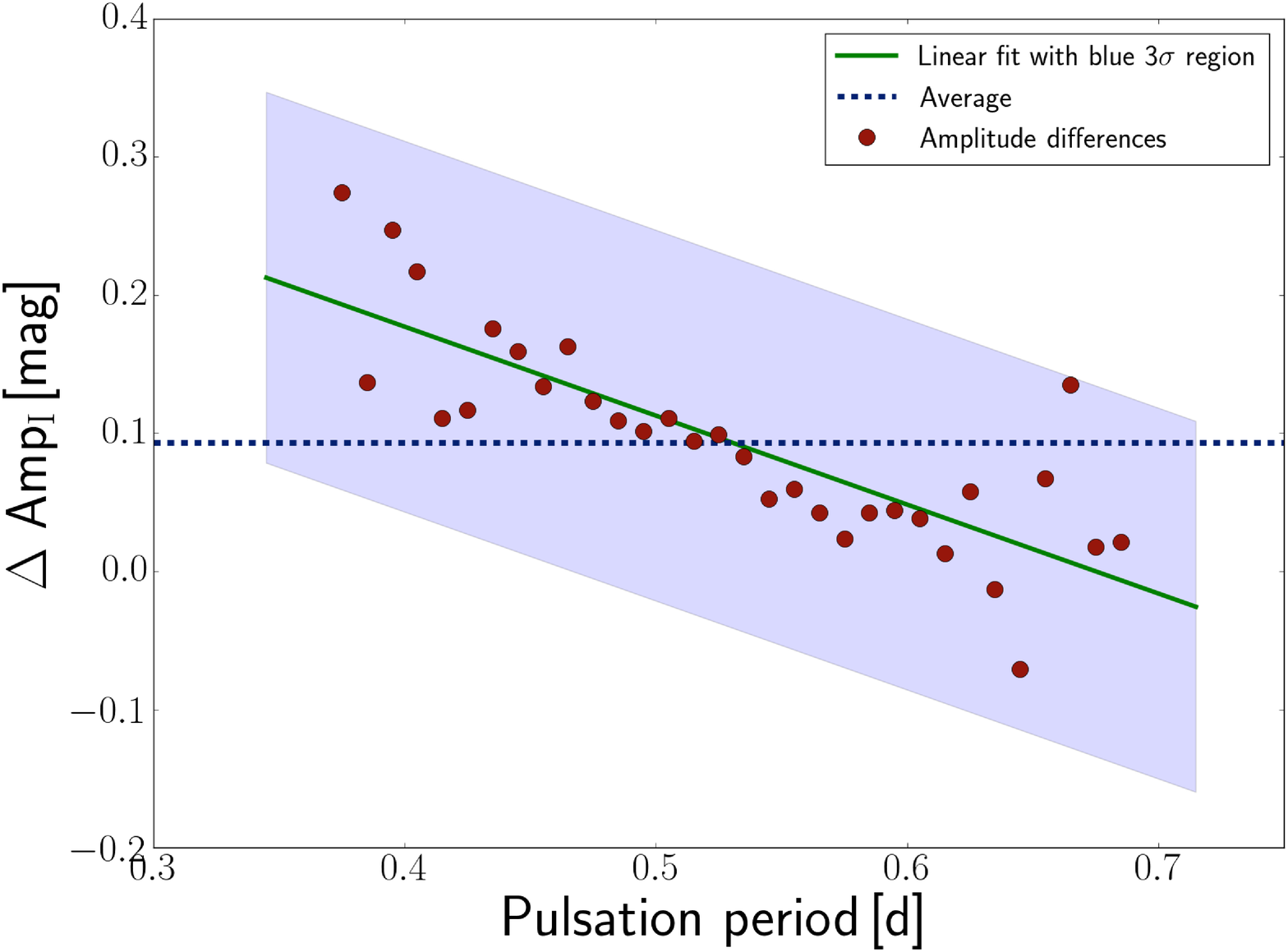}
    \caption{Diffrence in pulsation amplitudes of BL and single periodic stars computed in 32 period bins as $\Delta \rm{Amp_{I}=Amp^{nBL}_{medianI}-Amp^{BL}_{medianI}}$. The straight line shows the linear fit.}
    \label{Fig:AmpDifference}
\end{figure}

\subsection{Pulsation period}\label{Subsect:BLdistribution}

Fig.~\ref{Fig:IncidenceRatePeriodBins} shows the distribution of the BL stars population according to the pulsation period in 0.05-d bins. It is seen that the highest incidence rate of BL stars (about 48\,\%) appears in stars with pulsation period of 0.5\,d. Below 0.625\,d the average percentage is almost 44\,\%, while above this limit it drastically falls down to 20\,\%. The same behaviour was observed by \citet{smolec2005} in stars from GB and LMC; by \citet{szczygiel2007,skarka2014b} in stars from the Galactic field; and by \citet{jurcsik2011} in globular cluster M5. All these evidences suggest that the rare occurrence of the BL stars in RRLs with periods longer than 0.65\,d is real and has general plausibility regardless the investigated stellar system, its metal abundance and age. 


Because the pulsation period roughly relates to the mean density, the mass and size of an RRL star could be the factors that judge the occurrence of the modulation more than other physical parameters\footnote{Computed metallicity, intrinsic colour, and absolute magnitude are the same for BL and non-modulated stars, see the forthcoming sections.}. During the evolution the size and density of an RRL star gradually change and could be tuned to the `right' values allowing the modulation to rise. The mass-size parameter space for BL stars is probably somewhat limited preventing the modulation to be present in long-pulsation period RRL stars that have generally lower density.

The lack of modulated stars among long-pulsation period RRLs is the reason why the mean pulsation period of BL stars is shorter than for nBL stars: $0.533\pm 0.001$\,d vs. $0.564\pm 0.001$\,d. The difference of 0.031\,d well corresponds to what was found by \citet{jurcsik2011} and \citet{alcock2003}.

\subsection{Amplitudes of the light curves}\label{Subsect:Amplitudes}

Because the properties of the BL modulation have not been investigated within this study, we are unable to investigate light-curve amplitudes during the maximum BL phase, nor amplitudes of modulation-free light curves. We can only investigate the amplitudes of the mean light curves for BL stars. 

Although there are some exceptions from a general trend \citep[as was shown e.g. by][]{jurcsik2016}, it is generally accepted, that at given period, most of the light curves of modulated stars have amplitudes comparable to single-periodic RRLs only near their maximum-amplitude BL phase -- see fig. 3 in \citet{szeidl1988} and fig. 1 in \citet{jurcsik2016} that display amplitudes of RRLs in M3. Therefore, we can expect that at given period the mean light curves of BL stars will have smaller peak-to-peak amplitude (from minimum to maximum light) than their non-modulated counterparts.

This is to some extent apparent from the left panel of Fig.~\ref{Fig:Per-ampDistribution}, but it is obvious without a doubt from Fig.~\ref{Fig:AmpDifference}, where the difference between the mean amplitudes of the light curves of single-periodic RRLs and BL stars in 32 bins is plotted as a function of period\footnote{The most short- and long-period regions were omitted from the calculations because of low number of stars.}. To be more precise, we simply computed average value of amplitudes for non-modulated stars in a given period bin, the same we did for BL stars, and then we subtracted these two values. The amplitude difference shown in Fig.~\ref{Fig:AmpDifference} linearly decreases with increasing period as:
\begin{equation}\label{Eq:AmpDifference}
\Delta A_{I}=0.434(46)-0.643(86)P~~~~~~~~~~~~~~~~~~ \sigma=0.045.
\end{equation} 

Because the larger is the amplitude of the BL modulation, the lower usually is the total pulsation amplitude of a BL-star light curve, and because we see the largest difference in short-pulsation period RRLs, our result supports the finding by \citet{jurcsik2005}. They proposed that short-pulsation period BL stars can have both small and large modulation amplitudes, while long-period BL stars can have only small modulation amplitudes (resulting in smaller difference between BL and single-periodic RRLs).

\subsection{Rise time}\label{Subsect:RiseTime}

From the mean light curves we also estimated the rise time (RT), which is the phase difference between maximum and minimum light. Modulated stars have generally less skewed, more symmetric mean light curves with smaller amplitudes, thus, we can expect larger RT for BL stars. This is really observed: the mean {\it RT} of the BL stars is 0.2048(8), while nBL stars have the mean $RT=0.1690(4)$. The difference is clearly seen in the right-hand panels of Fig.~\ref{Fig:Per-ampDistribution}. The {\it RT} of BL stars is never less than approximately 0.16, while the majority of non-modulated RRLs has {\it RT} lower than 0.16. This finding could be used for the first, rough identification of BL stars in future classifications.

\subsection{Low-order Fourier coefficients}\label{Subsect:FourierCoefficients}

\begin{figure*}
	\includegraphics[width=\columnwidth]{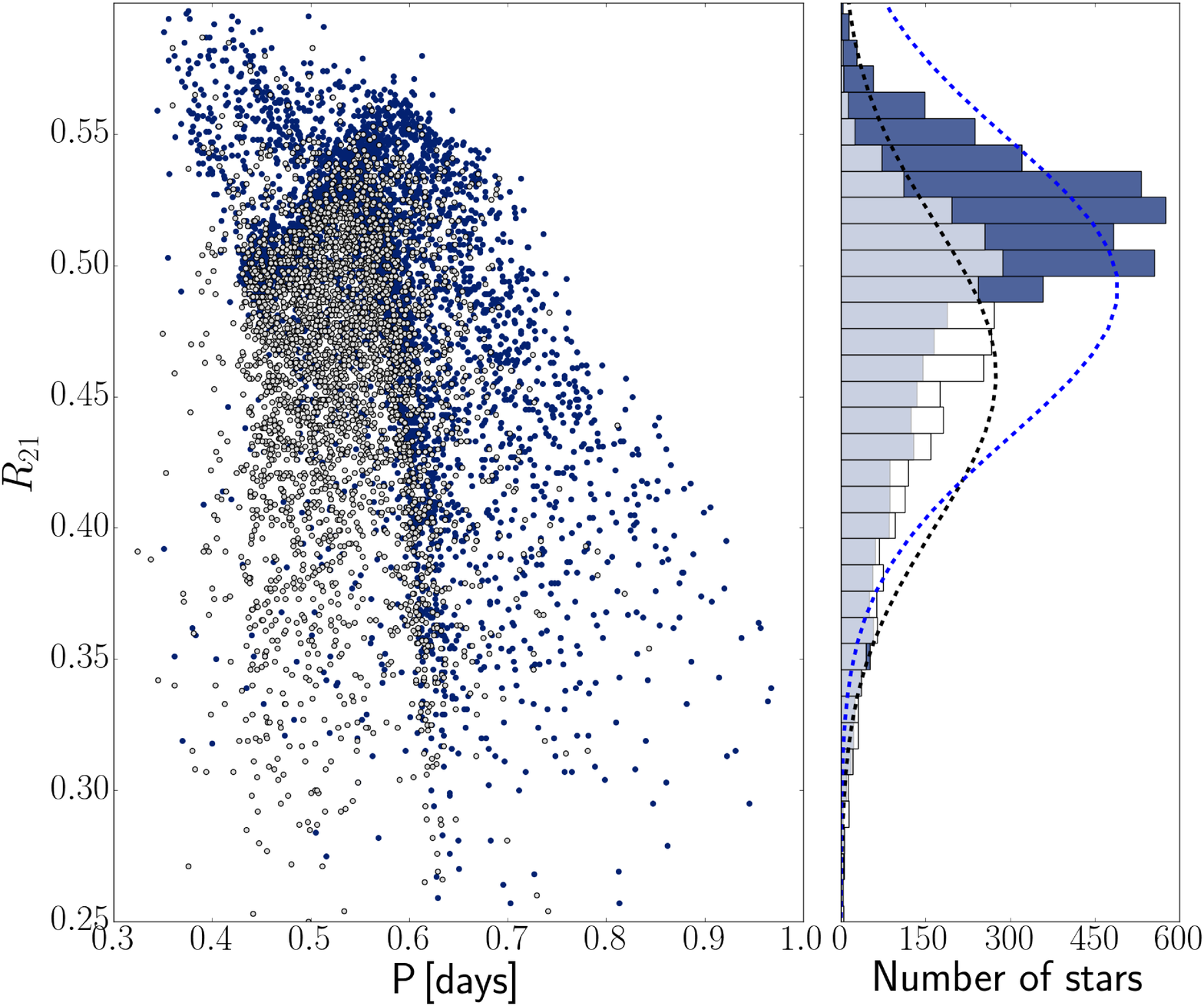}\includegraphics[width=\columnwidth]{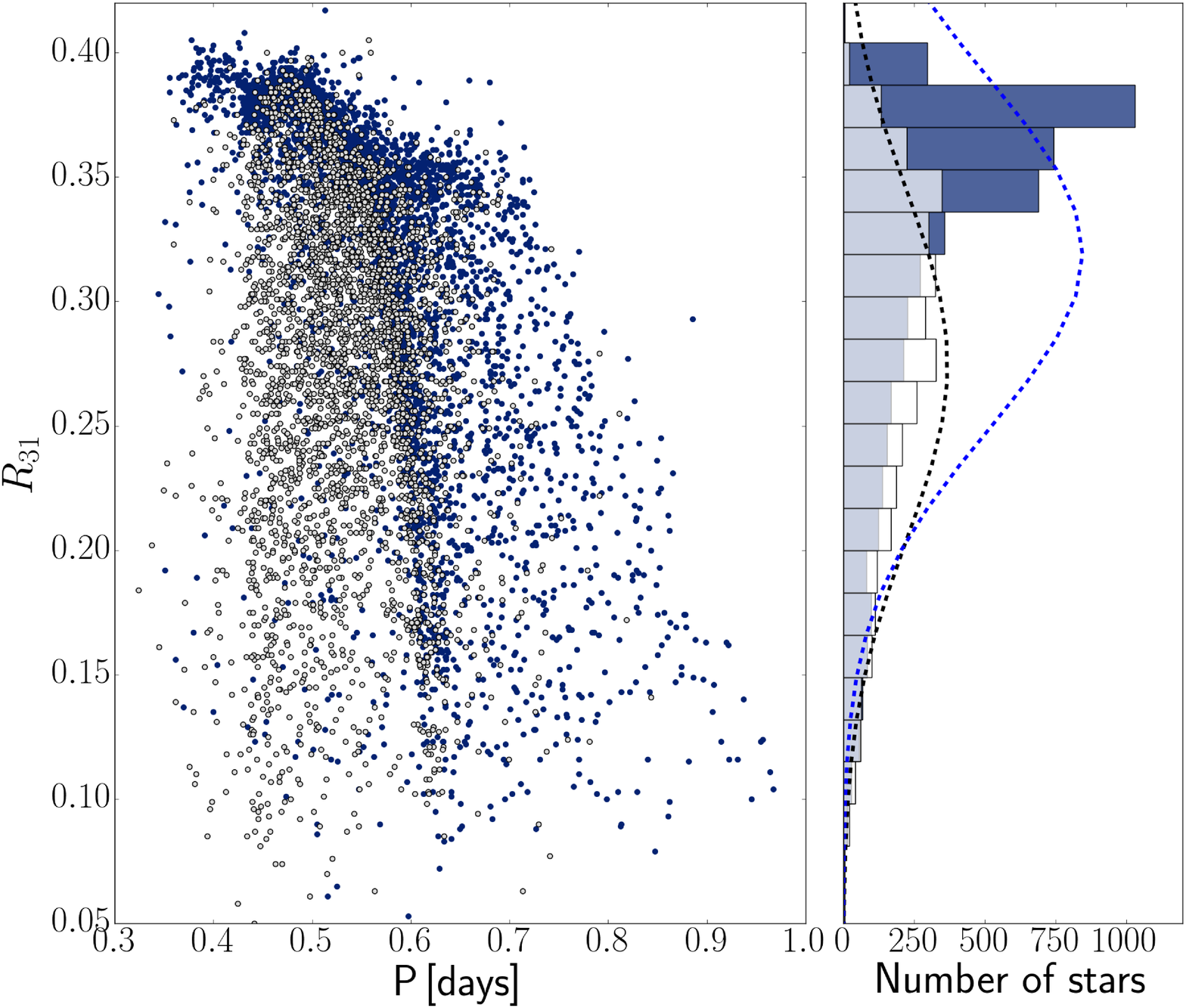}\\
	\includegraphics[width=\columnwidth]{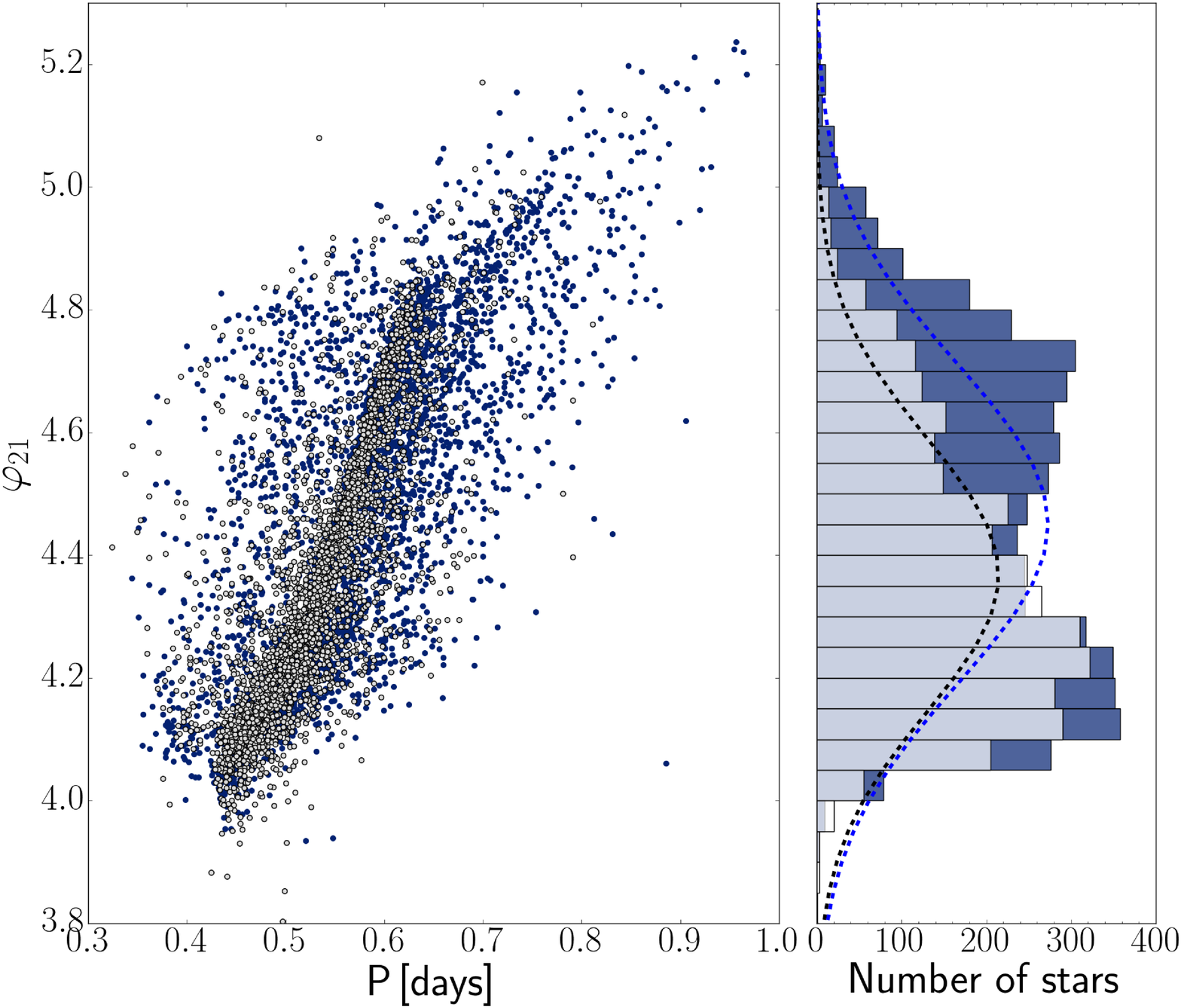}\includegraphics[width=\columnwidth]{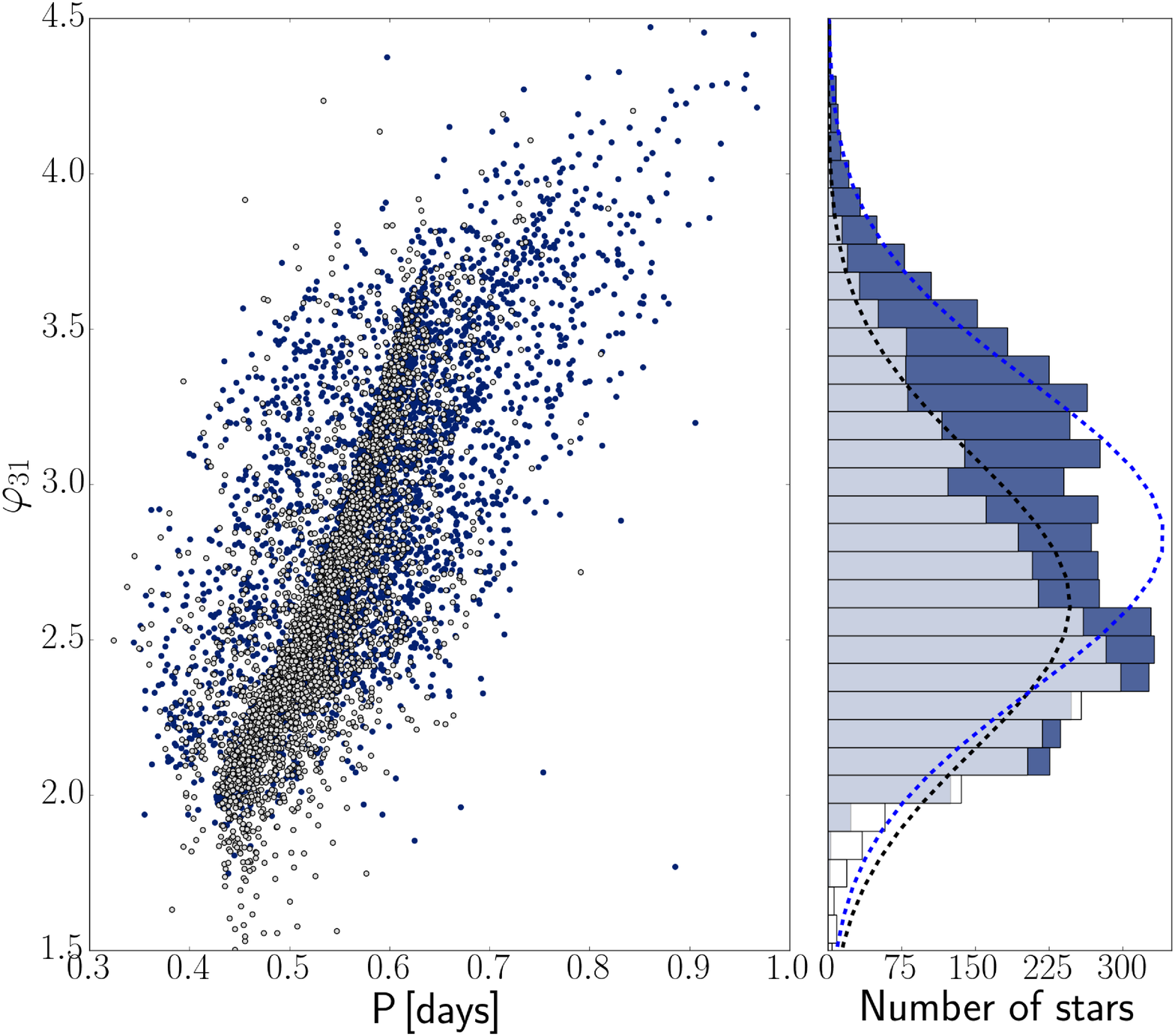}
    \caption{Low-order Fourier parameters. Normal distributions are shown with lines for comparison. Description of each plot is similar as in Fig.~\ref{Fig:Per-ampDistribution}.}
    \label{Fig:FourierParameters}
\end{figure*}

The mean values of Fourier parameters\footnote{Fourier parameters were introduced by \citet{simon1981} on Cepheid light curves. The parameters are defined as: $R_{i1}=A_{i}/A_{1}$ and $\phi_{i1}=\phi_{i}-i\phi_{1}$, where $A_{i}$ and $\phi_{i}$ are amplitudes and phases of the pulsation harmonics of the Fourier decomposition.} are systematically lower for BL stars than for stars with non-modulated light curve (see Table~\ref{Tab:Parameters}), which is naturally expected from the same reasons why RT is larger\footnote{We note again that we used exclusively parameters derived and provided by \citet{soszynski2014}.}. The lower values for BL stars are clearly apparent in all panels of Fig.~\ref{Fig:FourierParameters}. 

The Fourier parameters are correlated, which means that one parameter can be expressed as a linear combination of the others. The agreement between computed and observed values can serve as a criterion for testing whether a particular light curve is similar to the majority of RRL stars or not. The mean light curves of BL stars can,
especially when the strong modulation is present, significantly differ from those of single-periodic stars. \citet{jurcsik1996} introduced the so-called deviation parameter $D_{F}$, which reflects the normality of the light curve in $V$ passband. In addition, value of this parameter indicates whether their empirical equation for metallicity determination can be used for a particular star or not. 

We followed their approach and determined new linear interrelations between Fourier amplitudes and phases in $I$ filter. We used only 2260 non-modulated stars that are between 15 and 17\,mag, have more than 900 observations, and do not have pulsation periods between 0.49 and 0.51\,d to secure good quality light curves with homogeneous phase coverage. From the prewhintening procedure we have Fourier decomposition into ten sine harmonics that we used for this purpose\footnote{\citet{soszynski2014} do not provide amplitudes and phases of the Fourier decomposition.}. 

For each amplitude and phase up to the 6th order we performed linear fits with various combinations of other parameters. Because the Fourier coefficients are correlated, by using different combination of parameters one can get very similar results. Additional problem is that the more parameters are used, the better is the fit. Because classical decisive criteria (BIC, AIC) fail, one cannot easily decide what fit is the best. Therefore we used the full parameter fits employing all parameters. In Table~\ref{Tab:Interrelations} we give the solution. The most correlated components of the fits are highlighted with italics. 

\begin{figure*}
	\includegraphics[width=0.66\columnwidth]{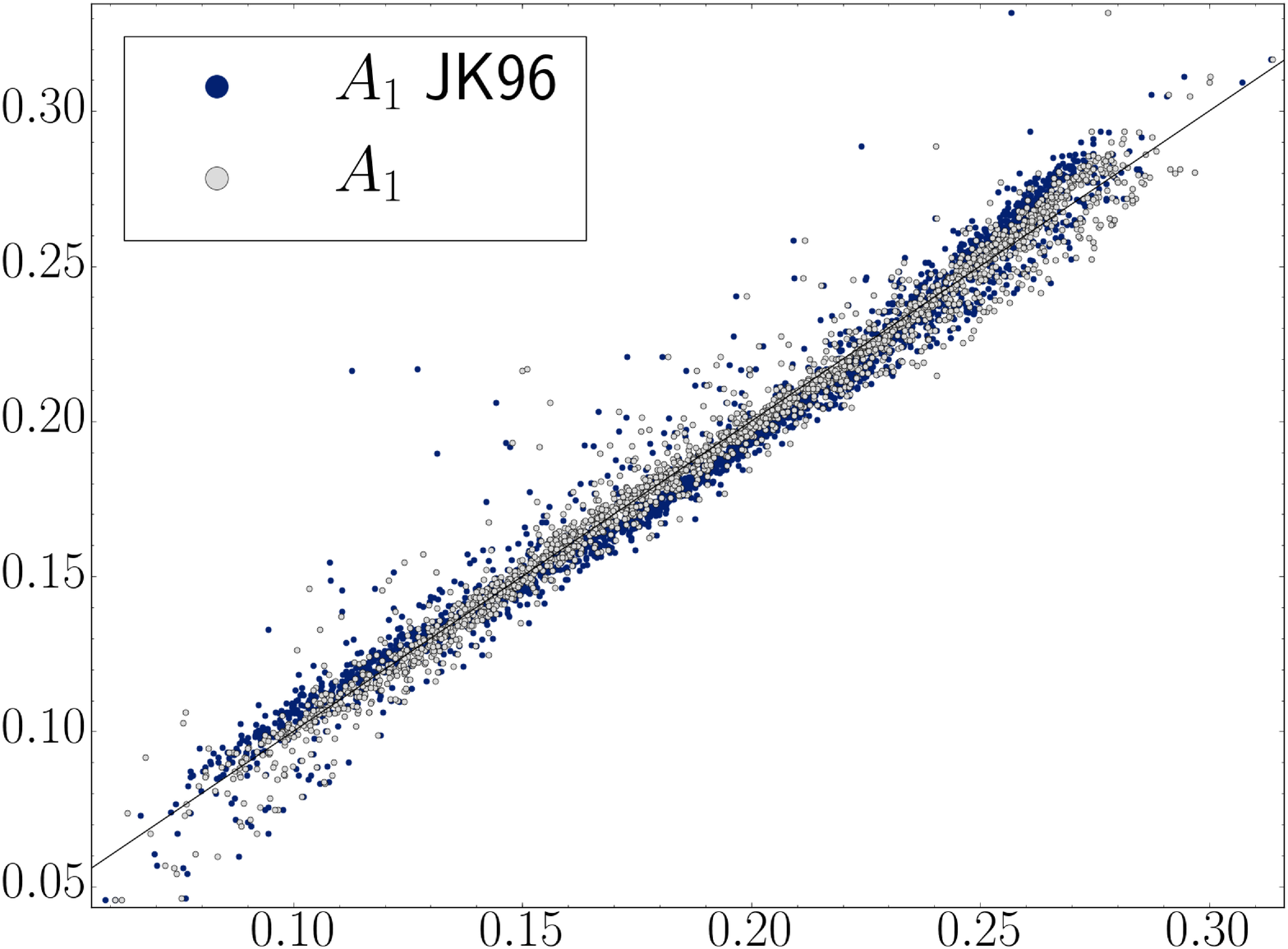}
	\includegraphics[width=0.66\columnwidth]{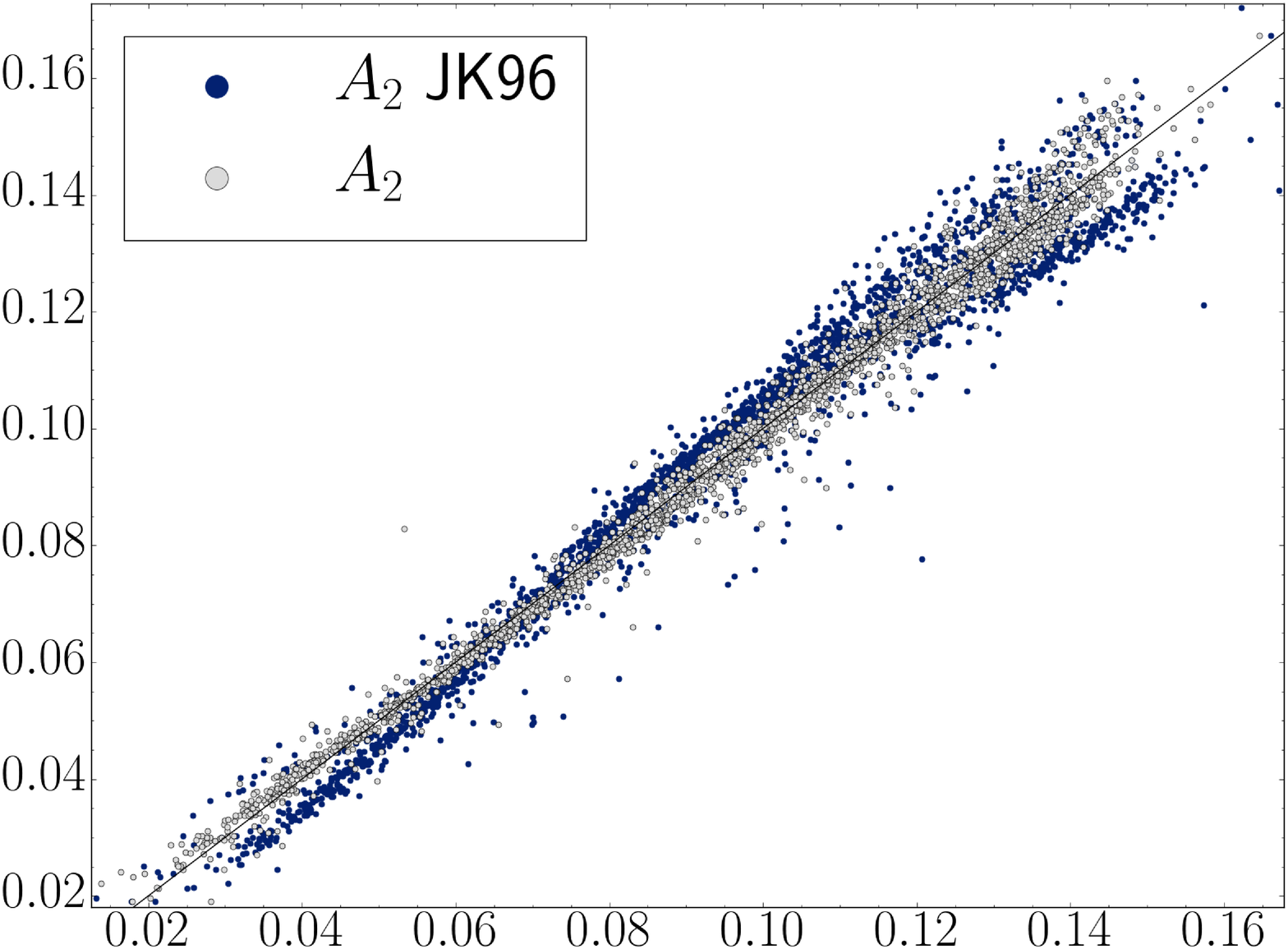}
	\includegraphics[width=0.66\columnwidth]{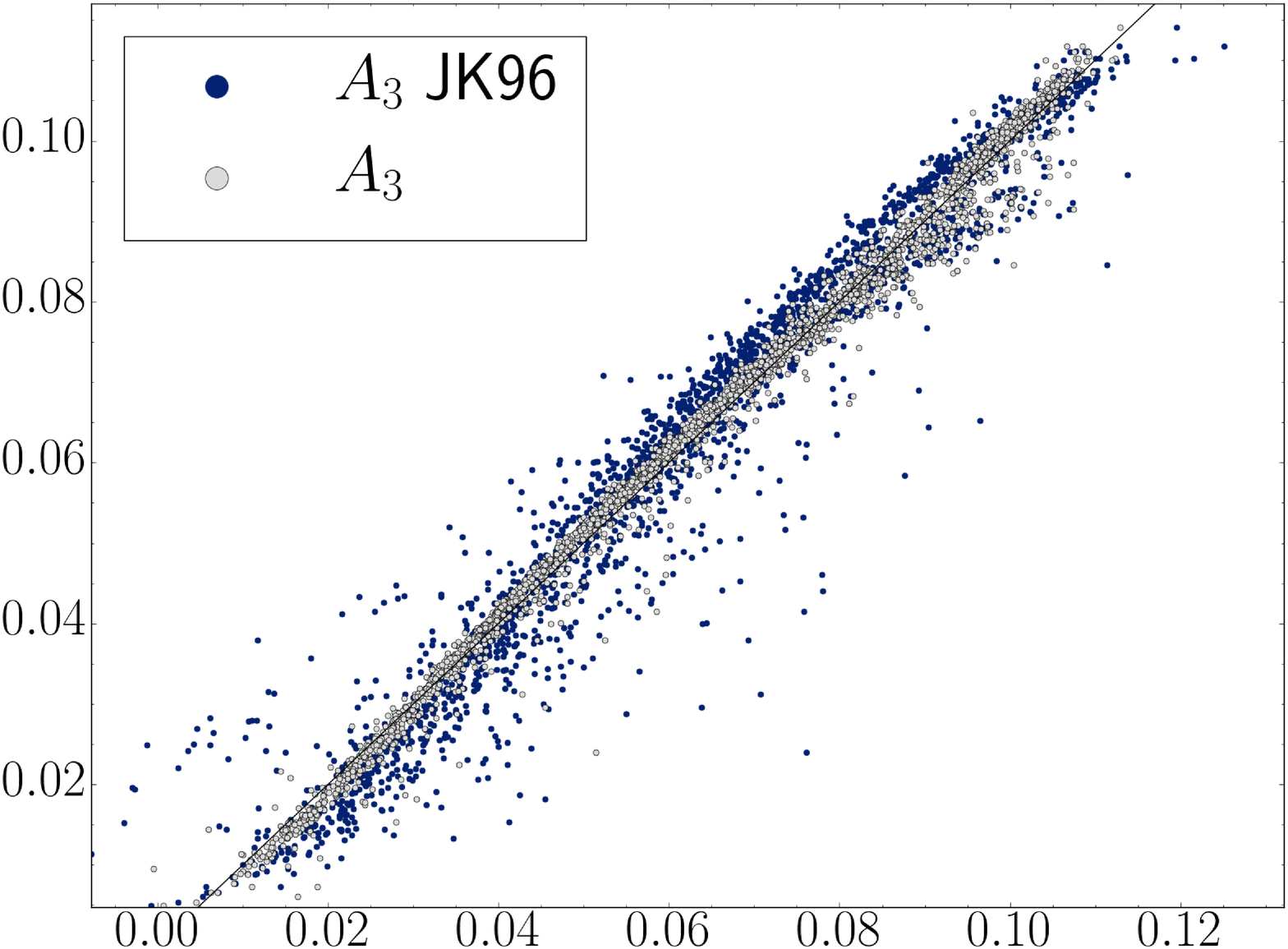}\\
	\includegraphics[width=0.66\columnwidth]{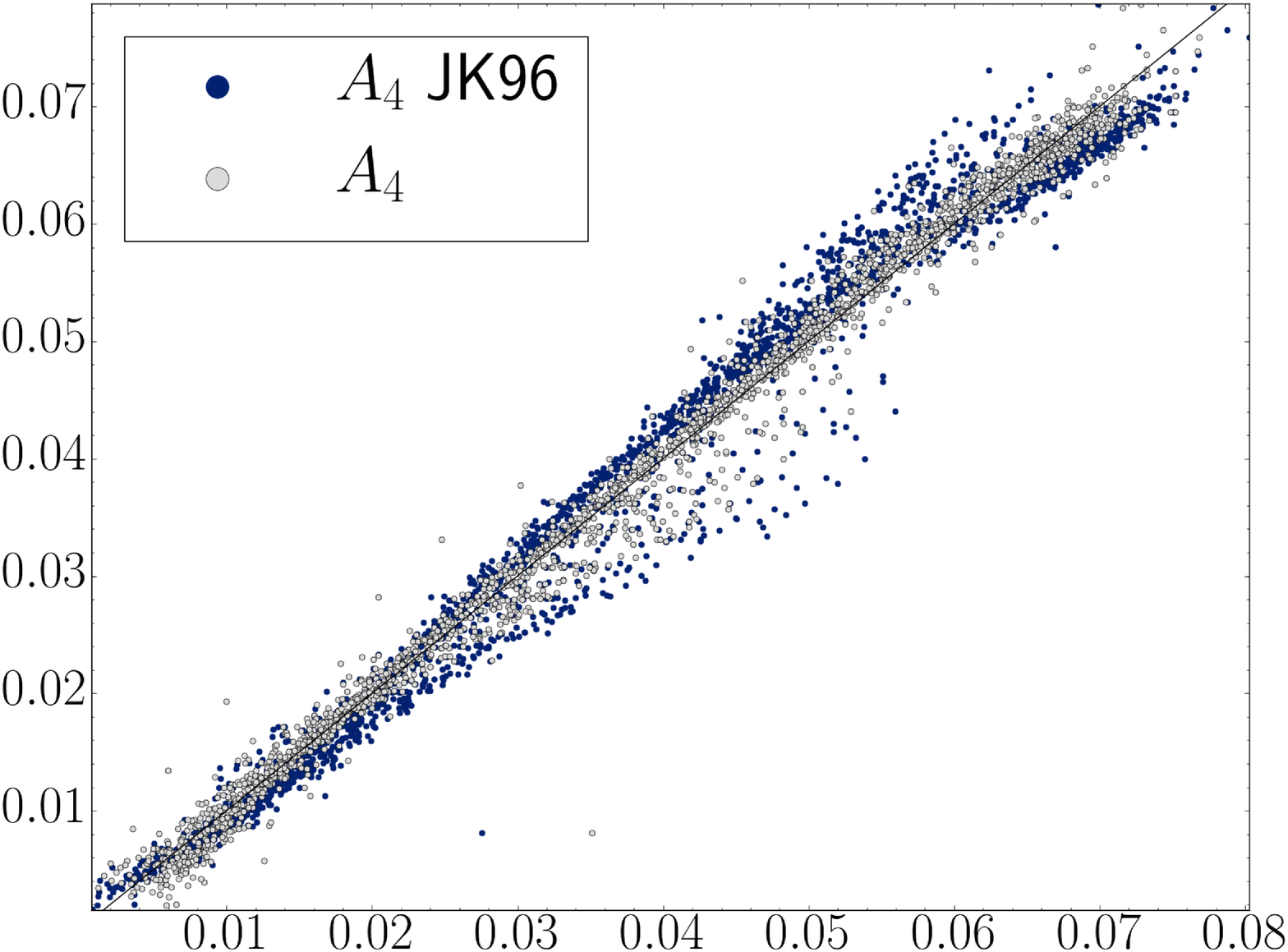}
	\includegraphics[width=0.66\columnwidth]{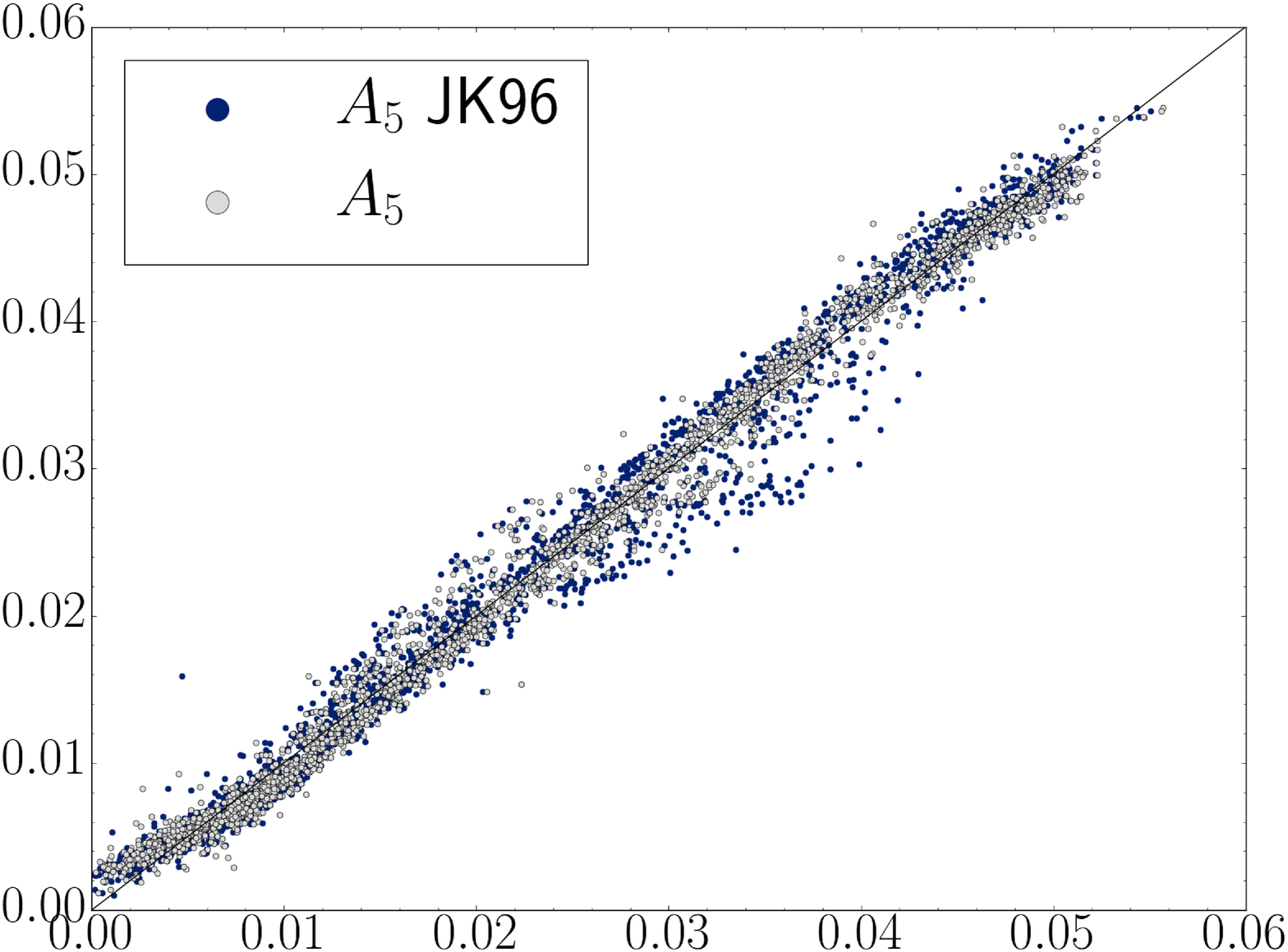}
	\includegraphics[width=0.66\columnwidth]{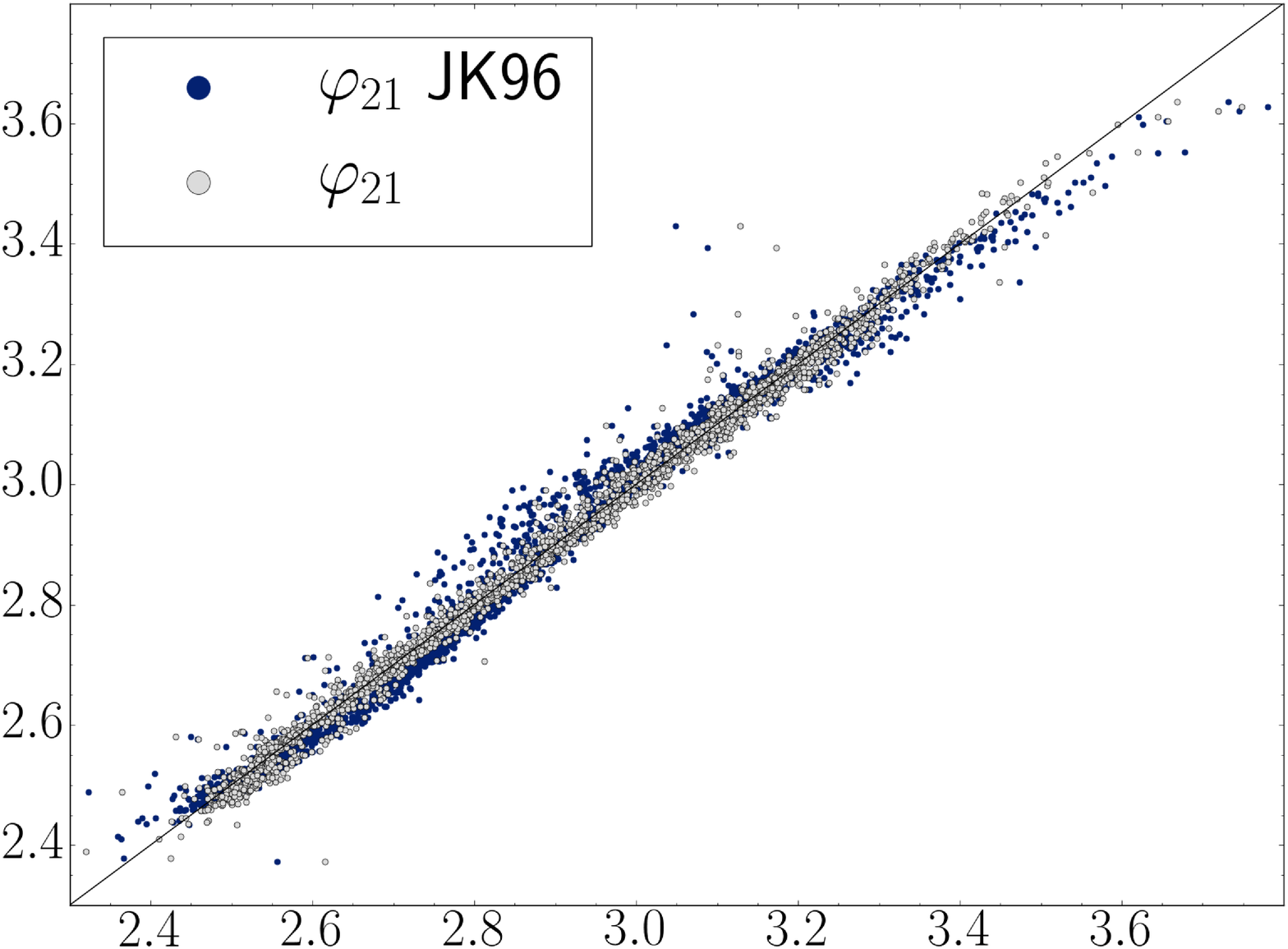}\\
	\includegraphics[width=0.66\columnwidth]{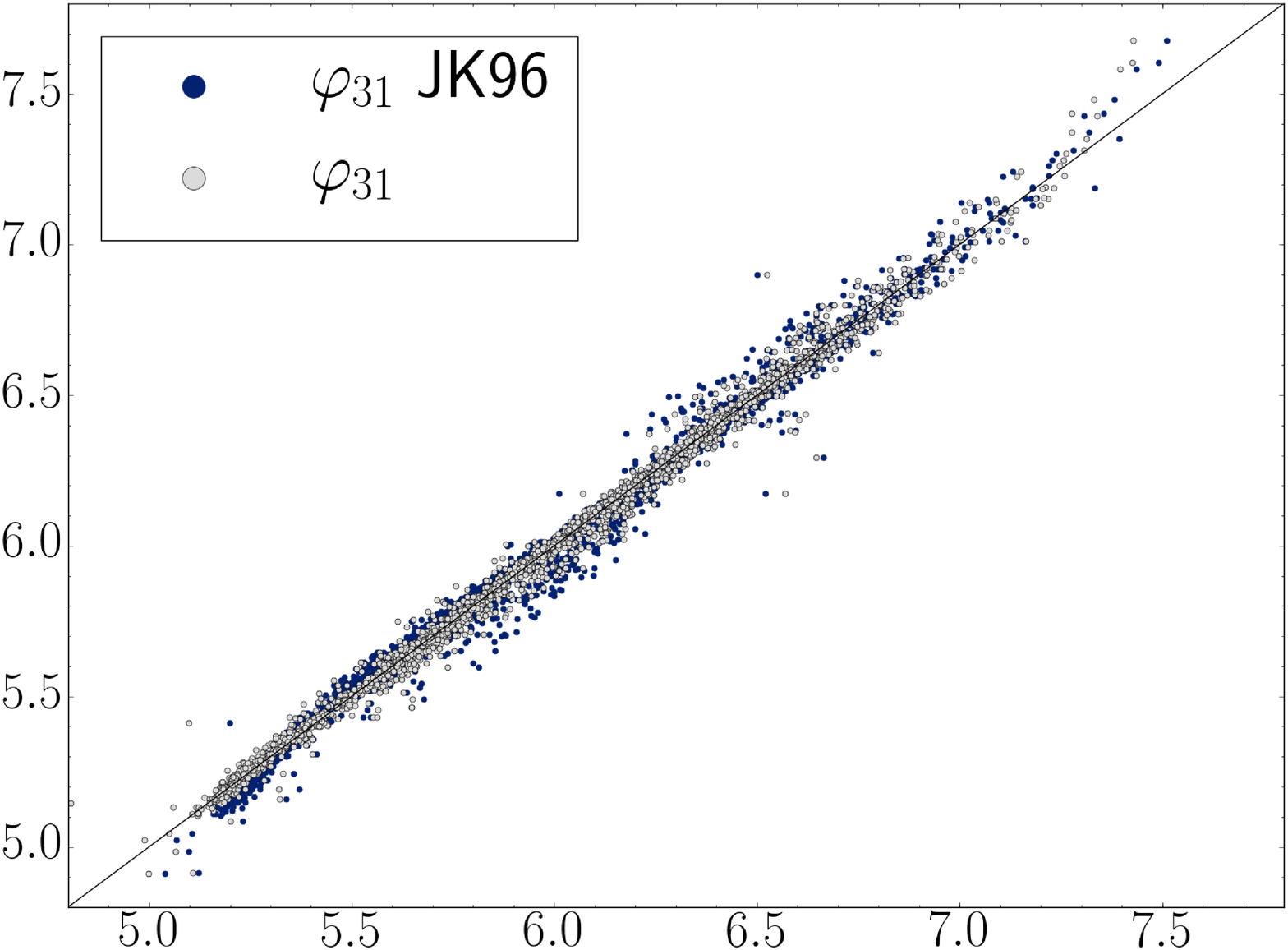}
	\includegraphics[width=0.66\columnwidth]{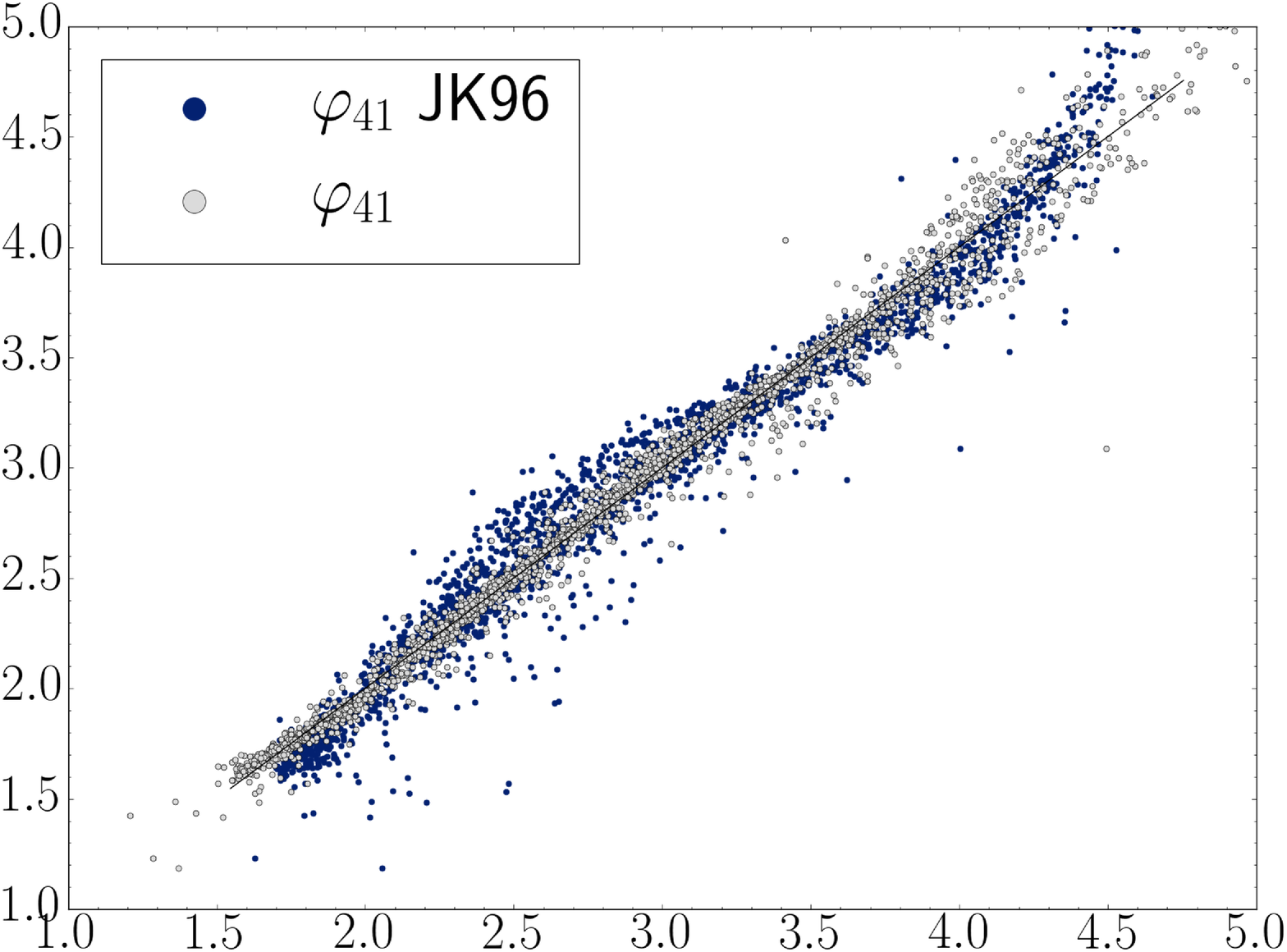}
	\includegraphics[width=0.66\columnwidth]{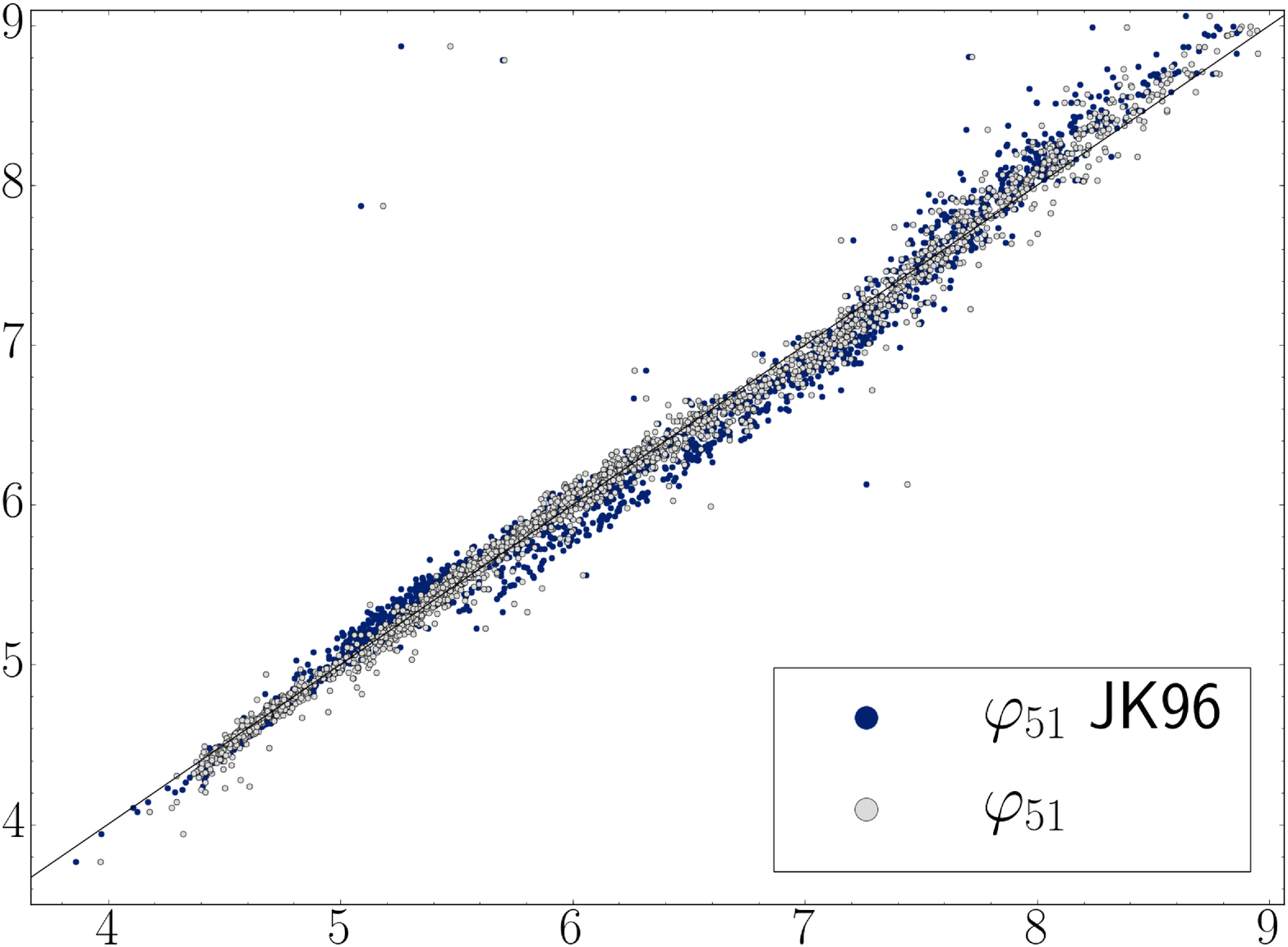}
	
    \caption{Calculated versus observed Fourier parameters. The straight line shows the 1:1 ratio.}
    \label{Fig:Interrelations}
\end{figure*}

Calculated versus observed parameters are shown in Fig.~\ref{Fig:Interrelations}. We also show how observed and calculated coefficients correlate in case when only particular amplitudes and phases used by \citet[][given in their table 6]{jurcsik1996} are used for the fit. 

The compatibility parameter $D_{F}$ is calculated according to eq. (6) from \citet{jurcsik1996} as $D_{F}=|F_{\rm obs}-F_{\rm calc}|/\sigma_{F}$, where  $F_{\rm obs}$ and $F_{\rm calc}$ rely to observational and calculated values of particular Fourier parameter $F$ according to fit in Table \ref{Tab:Interrelations} (with the standard deviation $\sigma_{F}$). The maximum of $\left\lbrace D_{F} \right\rbrace$ is $D_{m}$ that actually serves for the decision about applicability of the metallicity formula (in case that $D_{m}<3$).   

\setlength{\tabcolsep}{2pt}
\begin{table*}
\centering
\caption{Coefficients of the linear fit for the Fourier coefficients.}
\label{Tab:Interrelations}
\footnotesize
\begin{tabular}{lccccccccccccc}
\hline
&Const&$A_{1}$&$A_{2}$&$A_{3}$&$A_{4}$&$A_{5}$&$A_{6}$&$\phi_{21}$&$\phi_{31}$&$\phi_{41}$&$\phi_{51}$&$\phi_{61}$&$\sigma$\\ \hline
$A_{1}$&-0.08(1)&-&{\it 1.23(4)}&0.62(6)&-1.07(7)&1.5(1)&0.5(1)&0.045(6)&-0.003(4)&-0.008(2)&0.0027(6)&0.0014(4)&0.008\\
$A_{2}$&-0.082(5)&{\it 0.243(8)}&-&0.58(3)&1.08(3)&-1.18(5)&0.14(5)&0.022(3)&0.005(2)&-0.0056(6)&0.0006(3)&-0.0009(2)&0.003\\
$A_{3}$&0.050(4)&0.36(2)&0.076(8)&-&{\it 0.06(3)}&0.59(4)&-0.23(4)&0.009(2)&-0.011(2)&0.00005(50)&-0.0023(2)&0.0013(4)&0.003\\
$A_{4}$&0.059(3)&-0.075(6)&0.38(1)&{\it 0.03(2)}&-&0.8(3)&-0.26(3)&-0.024(2)&-0.001(1)&0.0060(4)&0.0004(2)&-0.0006(1)&0.002\\
$A_{5}$&-0.035(2)&0.049(4)&-0.203(8)&0.16(1)&0.4(1)&-&{\it 0.67(1)}&-0.016(1)&0.0150(6)&-0.0038(3)&0.0001(1)&-0.0005(8)&0.002\\
$\phi_{21}$& 0.22(4)&0.5(7)&1.3(2)&0.8(2)&-4.1(3)&-5.5(4)&5.0(4)&-&{\it 0.426(9)}&0.038(5)&-0.007(2)&-0.005(2)&0.026\\
$\phi_{31}$&1.94(6)&-0.1(1)&0.8(3)&-2.8(4)&-0.6(5)&14.3(6)&-14.7(5)&{\it 1.16(3)}&-&0.022(8)&0.005(4)&0.013(3)&0.043\\
$\phi_{41}$&-6.34(9)&-1.5(3)&-5.2(7)&0.1(1)&15(1)&-20(2)&25(1)&0.58(8)&{\it 1.24(5)}&-&0.073(8)&-0.010(6)&0.102\\
$\phi_{51}$&4.4(4)&3.7(8)&4(1)&-26(2)&8(3)&5(4)&-1(3)&-0.9(2)&0.2(1)&{\it 0.56(6)}&-&{\it 0.45(1)}&0.281\\ \hline
\end{tabular}
\end{table*}  

\begin{figure*}
	\includegraphics[width=\columnwidth]{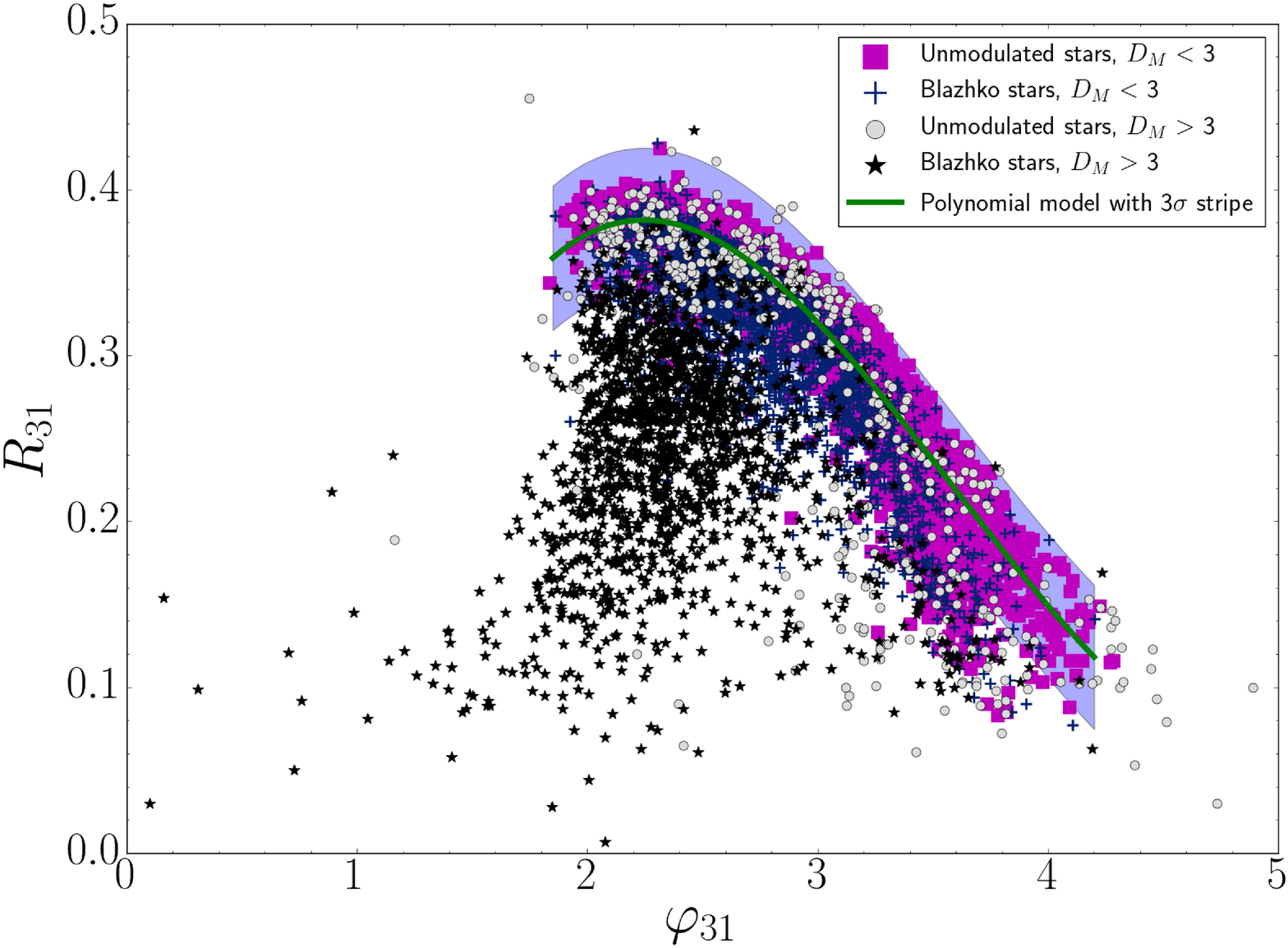}\includegraphics[width=\columnwidth]{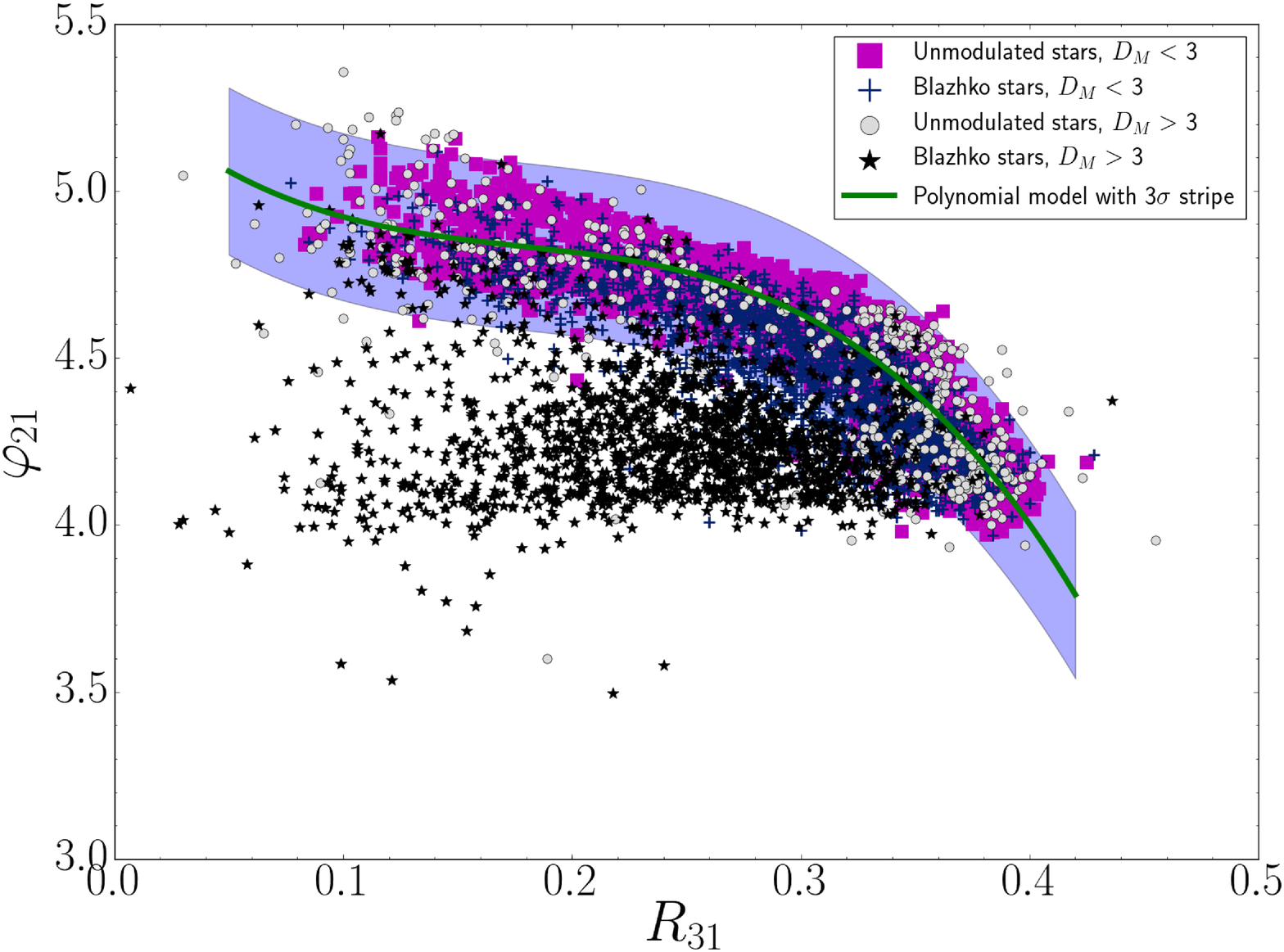}
    \caption{Relation of $R_{31}$ with low order phase parameters. The polynomial fits (equations \ref{Eq:Diff1} and \ref{Eq:Diff2} are shown with the solid line together with $3\sigma$ limits.)}
    \label{Fig:FourierParametersPopFits}
\end{figure*}

As was pointed out by \citet{cacciari2005}, the $D_{m}$ parameter cannot be effectively used for distinguishing between non-modulated and BL stars as was originally suggested by \citet{jurcsik1996}, because lot of BL stars have $D_{m}<3$. However, among stars with $D_{m}>3$ almost all stars are modulated! It is clearly seen in Fig.~\ref{Fig:FourierParametersPopFits}. While umodulated stars with all possible $D_{m}$ and BL stars with $D_{m}<3$ follow the same trend, BL stars with $D_{m}>3$ are located in completely different area in the $R_{31}$ vs. $\phi_{31}$ and $\phi_{21}$ vs. $R_{31}$ planes. 

This led us to define new limit for BL stars that is alternative for the $D_{m}$ criterion. The main streams in the $R_{31}$ vs. $\phi_{31}$ and $\phi_{21}$ vs. $R_{31}$ planes can be described by the polynomials in the following form:
\begin{equation}\label{Eq:Diff1}
\begin{array}{lllll}
\phi_{21}&=5.31&-6.53R_{31}&+32.4R^{2}_{31}&-60.7R^{3}_{31},\\
& \pm 0.04 &\pm 0.53  &\pm 2.1 &\pm 2.7, \\ 
\end{array}
\end{equation}
\begin{equation}\label{Eq:Diff2}
\begin{array}{lllll}
R_{31}&=-0.66&+1.1\phi_{31}&-0.36\phi^{2}_{31}&+0.033\phi^{3}_{31}, \\
& \pm 0.03 &\pm 0.03  &\pm 0.01 &\pm 0.001
\end{array}
\end{equation} 

The dispersion of eq. \ref{Eq:Diff1} and eq. \ref{Eq:Diff2} is $\sigma =0.083$\,rad and $\sigma = 0.015$, respectively. As it is seen from the Fig.~\ref{Fig:FourierParametersPopFits} and from the Table \ref{Tab:Fits}, the population of stars below $3\sigma$ from the fit consists from 94\,\% of BL stars. In other words, our equations could help to identify new BL stars in large sample of stars with 94\% probability. We performed detail analysis of 360 randomly chosen non-modulated stars below the $1\sigma$ region and identified only 8 additional stars that are in fact modulated, which is 2.2\,\% of all stars in this region\footnote{These stars are marked with `h' in Table~\ref{Tab:ListOfStars}.}. This percentage could serve as the estimation of our failure in searching for the BL effect due to human factor and used methods. It is worth to note that these 8 stars are somewhat peculiar because the side peaks at higher harmonics have larger amplitudes than side peaks around $f_{0}$ and $2f_{0}$\footnote{In some of the stars the side peaks around $f_{0}$ and $2f_{0}$ were even undetectable.}, where we originally searched for the peaks.

\begin{table}
	\centering
	\caption{Occurence rate of the BL stars and the total number of stars in different distances below the fits.}
	\label{Tab:Fits}
	\begin{tabular}{lccc} 
		\hline
		n$\sigma$ & $R_{31}$ vs. $\phi_{21}$ & $R_{31}$ vs. $\phi_{31}$ \\
		\hline
		1 & 73.9\,\% (3231) & 71.5\,\% (3695) \\
		2 & 89.0\,\% (2111) & 78.7\,\% (2855) \\
		3 & 94.4\,\% (1602) & 83.6\,\% (2281) \\
		\hline
	\end{tabular}
\end{table}

\subsection{Metallicity}\label{Metallicity}

\begin{figure}
	\includegraphics[width=\columnwidth]{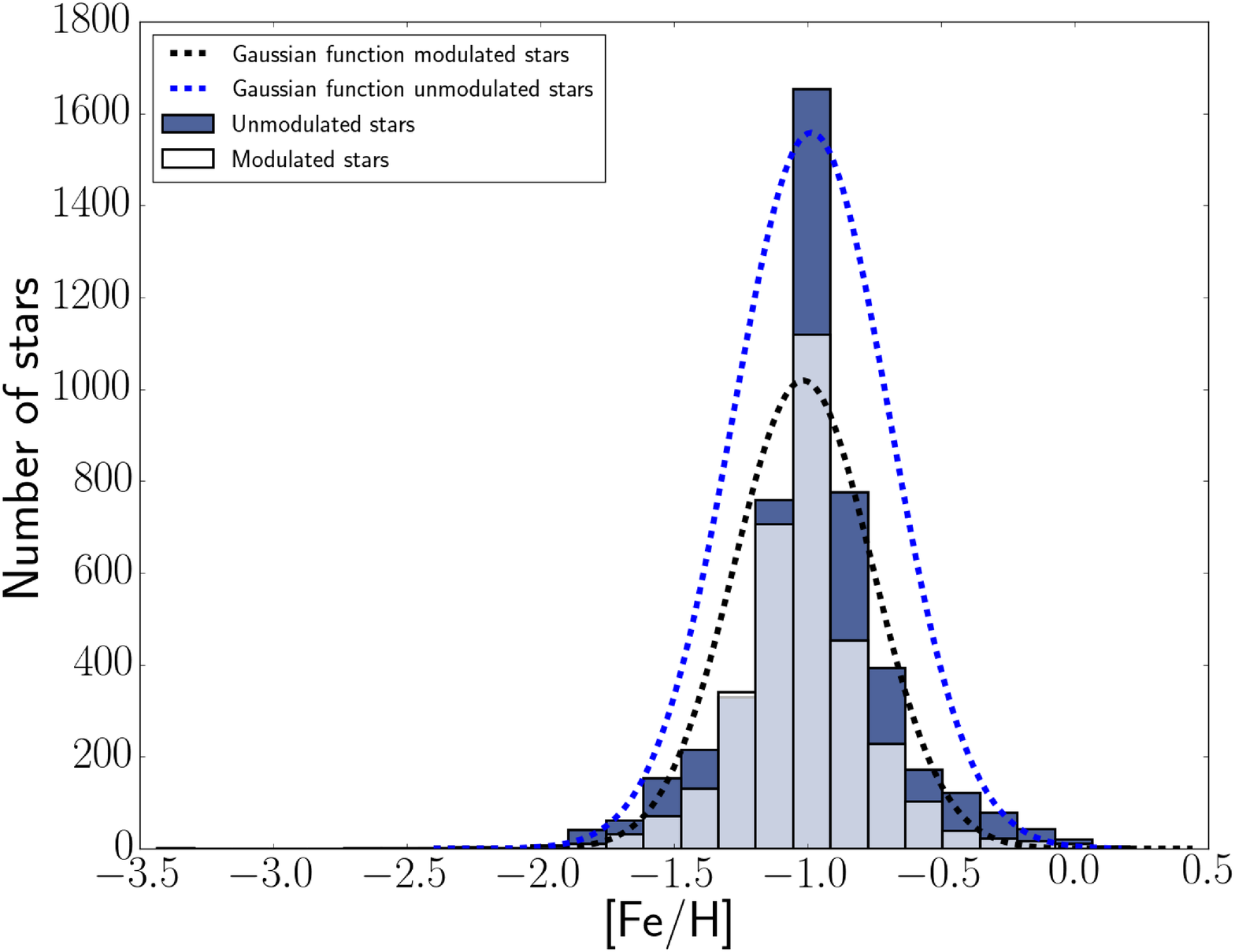}
    \caption{Distribution of metallicity for BL and single-periodic stars with $D_{m}<3$. Gaussian distributions with the same parameters as have the corresponding populations are shown with lines for comparison.}
    \label{Fig:MetalDistribution}
\end{figure}

For stars that meet the $D_{m}<3$ criterion, we can apply the empirical relation by \citet{jurcsik1996} modified for $I$ filter by \citet[][his equation 2]{smolec2005} and investigate differences in metallicity for BL and non-modulated stars. The results, that are shown in Fig.~\ref{Fig:MetalDistribution} and in Table \ref{Tab:Parameters}, suggest that there is no difference in metallicity between the two groups of stars at all. Thus, the (non)presence of modulation in a particular star depends more likely on other physical parameters (such as, for example, the size and density) than on the metal abundance. Similar results for the GB were already reported by \citet{smolec2005}, and by \citet{skarka2014b} for Galactic field stars.

\subsection{Intrinsic colours and absolute magnitudes}\label{Subsect:ColoursLuminosity}

\begin{figure}
	\includegraphics[width=\columnwidth]{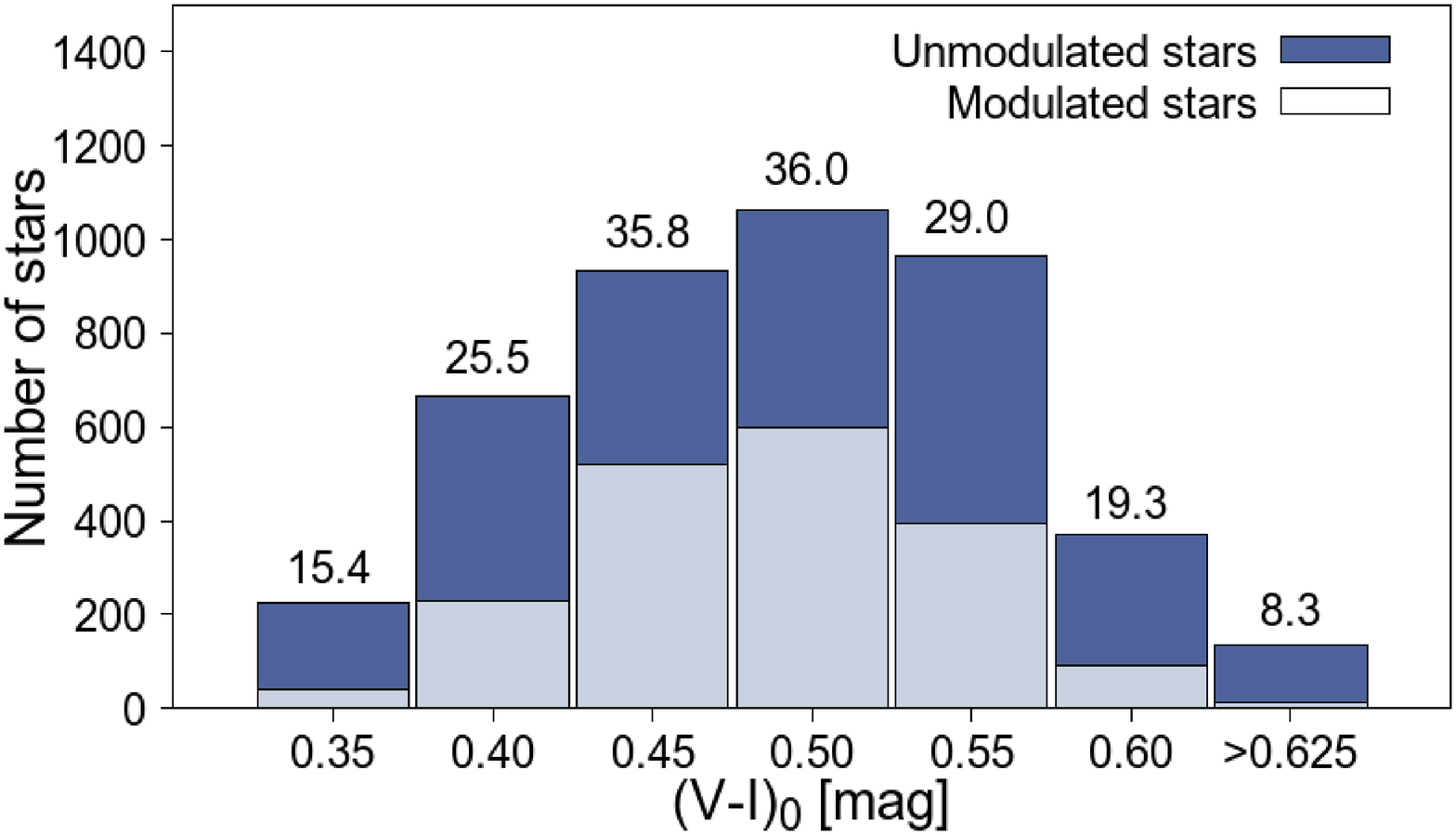}
    \caption{Distribution of $(V-I)_{0}$ for BL and single-periodic stars with $D_{m}<3$. The percentage of BL stars is shown above each bin.}
    \label{Fig:ColourDistribution}
\end{figure}

Following \citet{pietrukowicz2015}, we computed the intrinsic colour $(V-I)_{0}$ using theoretical calibrations by \citet{catelan2004} based on calculations of synthetic horizontal branches. This approach simplifies the problem and allows for computing the intrinsic colour without knowing the extinction. The used formula is:
\begin{equation}\label{Eq:Colour}
(V-I)_{0}=1.817+1.132\log P +0.677\log Z +0.108(\log Z)^{2},
\end{equation}
where $\log Z$ is calculated as
\begin{equation}
\log Z=\rm{[Fe/H]}-1.765.
\end{equation}
BL and non-modulated stars with $D_{m}<3$ have the same average $(V-I)_{0}$ suggesting that there is no correlation between BL effect and colour in the GB (see Table \ref{Tab:Parameters} and Fig.~\ref{Fig:ColourDistribution}). 

Absolute magnitudes in various pass bands could also be computed using the calibrations provided by \citet{catelan2004}. In Table \ref{Tab:Parameters} we give the absolute magnitude in $I$ for both groups calculated as:
\begin{equation}\label{Eq:Imag}
M_{I}=0.417-1.132\log P + 0.205\log Z,
\end{equation}
\citep[eq. 3 in][]{catelan2004}. The mean values in Table \ref{Tab:Parameters} show that modulated and nBL stars with $D_{m}<3$ have the same average photometric absolute magnitude/luminosity.

\subsection{Spatial distribution}\label{Subsect:SpatialDistribution}

\begin{figure*}
	\includegraphics[width=\columnwidth]{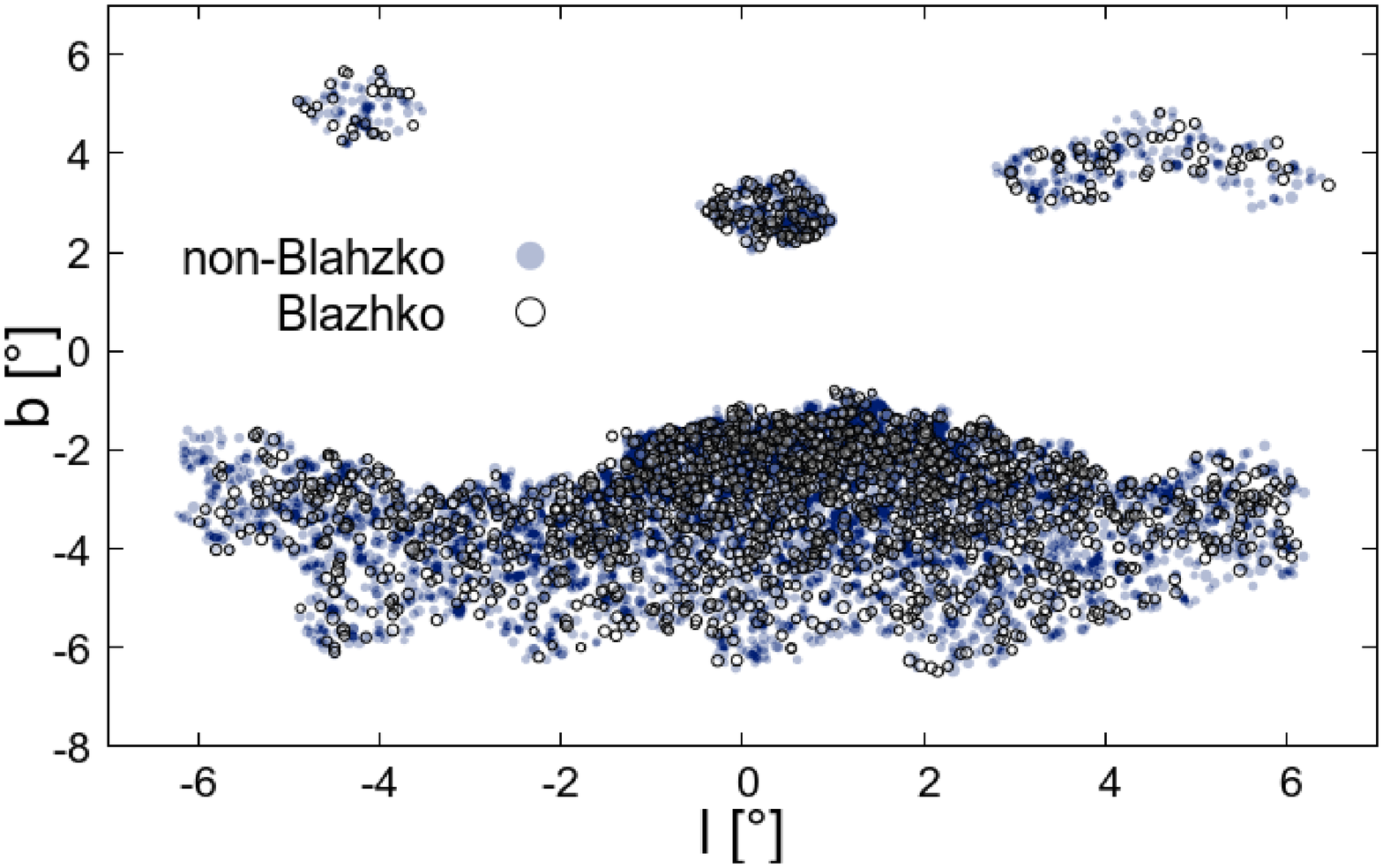}\includegraphics[width=\columnwidth]{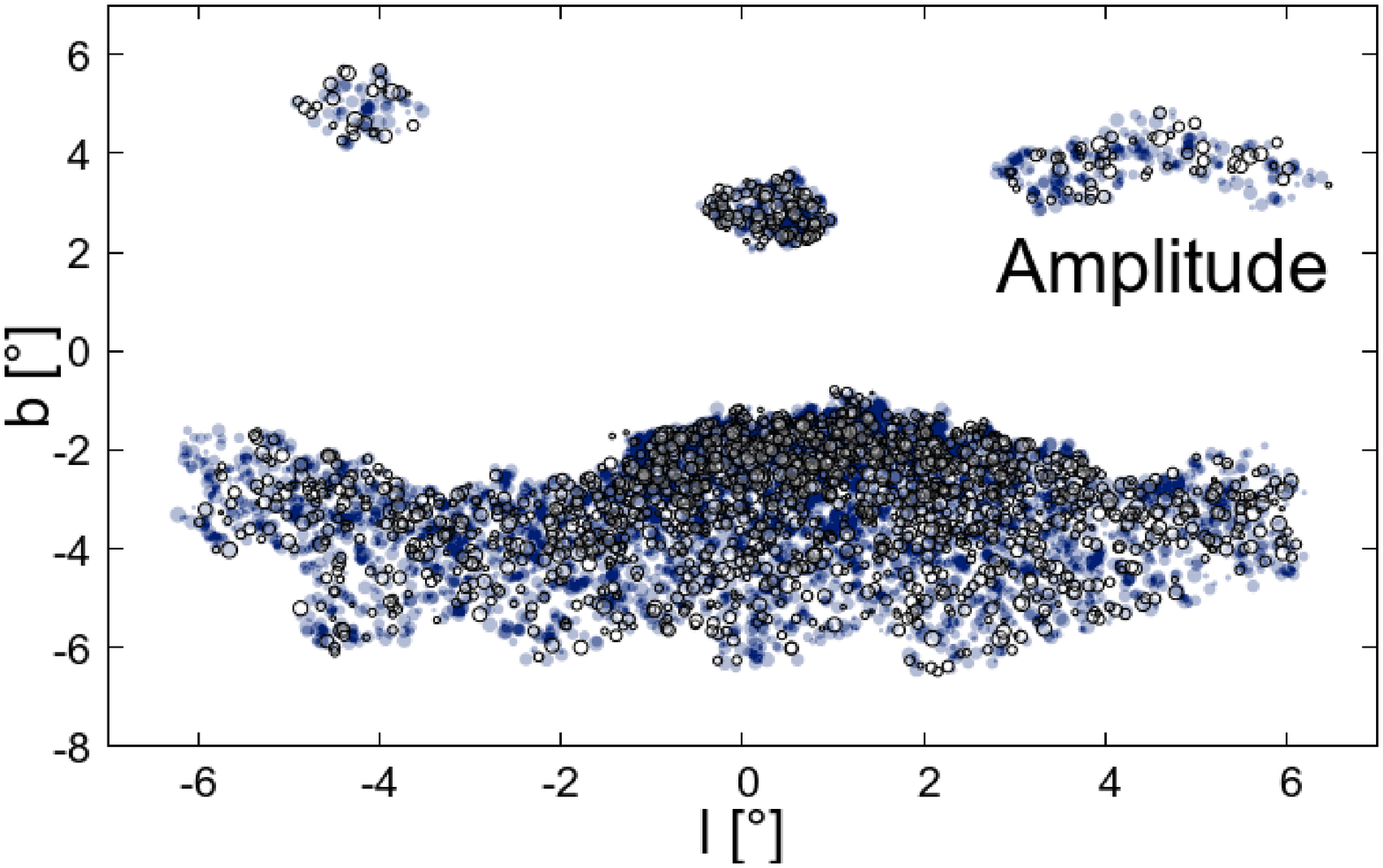}\\
	\includegraphics[width=\columnwidth]{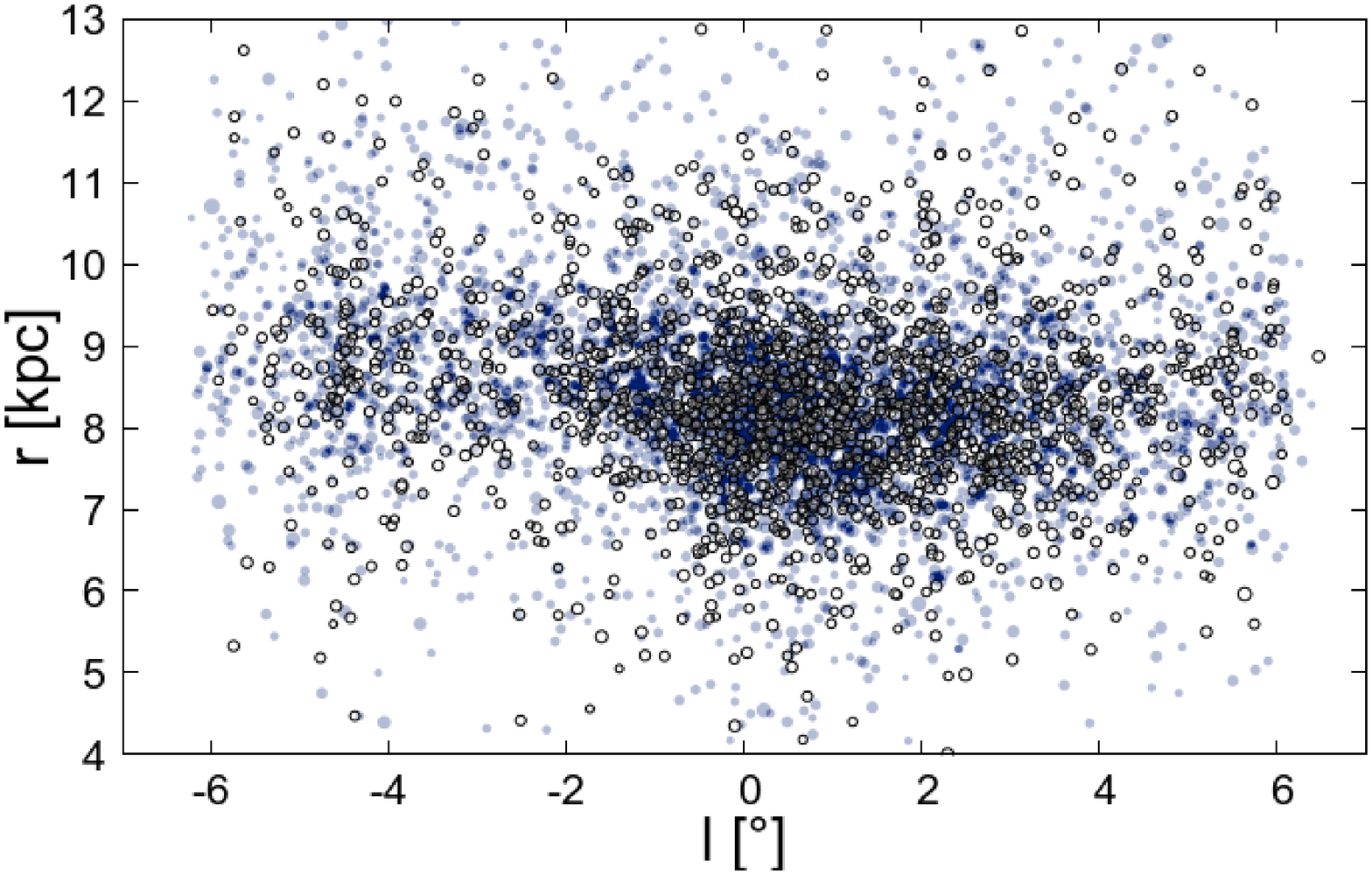}\includegraphics[width=\columnwidth]{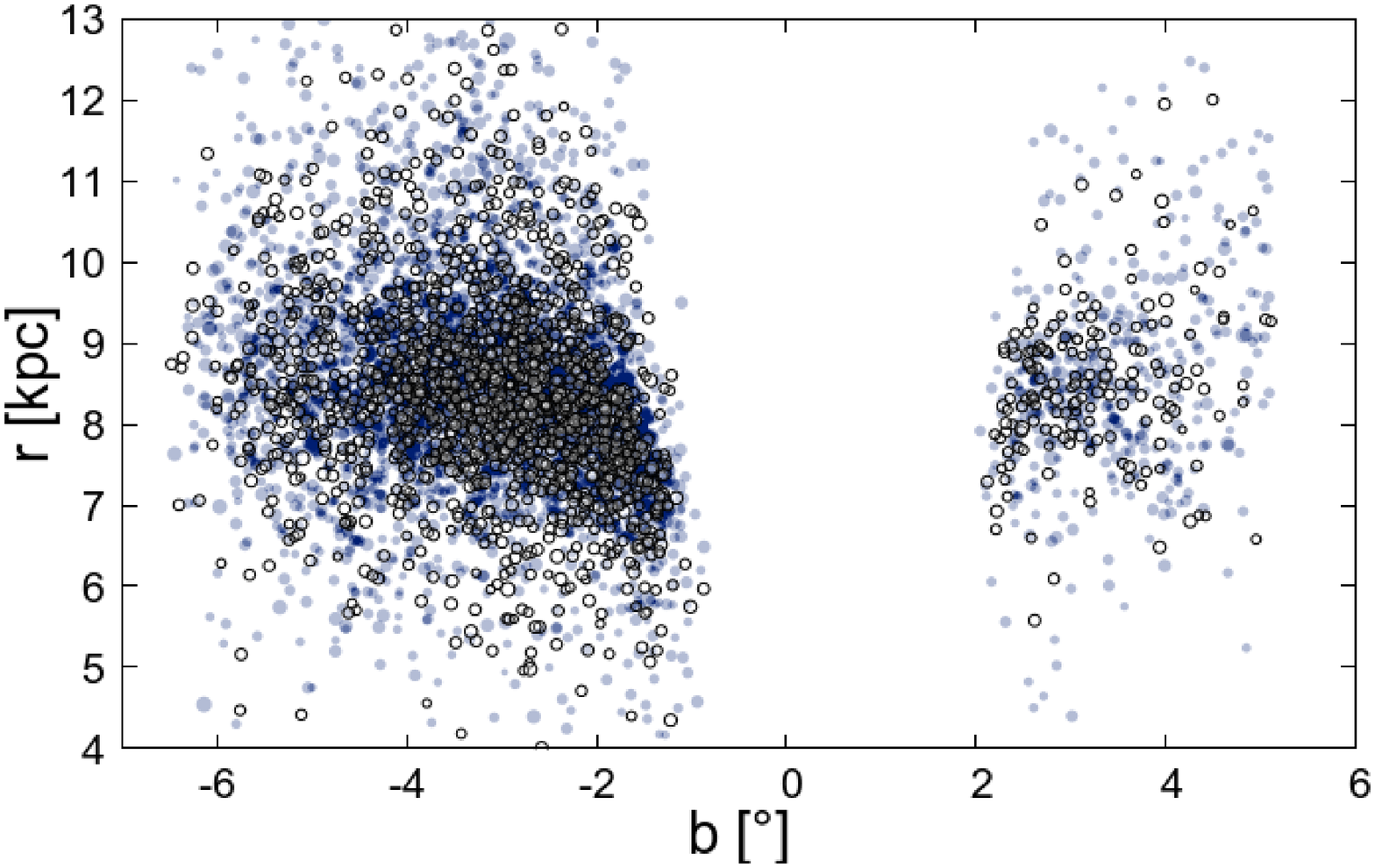}\\
    \caption{Spatial distribution of sample BL (white circles) and non-modulated stars (blue circles). The larger are the points, the longer is the pulsation period. In the top right-hand panel, the size of circles is proportional to the pulsation amplitude, for comparison.}
    \label{Fig:SpatialDistribution}
\end{figure*}

The distribution of modulated stars seems to correlate with the spatial distribution of stars with nBL light curve with no preference towards any direction (Fig.~\ref{Fig:SpatialDistribution}). To our knowledge, this is for the first time when the investigation in spatial distribution of BL stars is possible. In calculation of the distance to the stars we followed exactly the same procedure as described in \citet{pietrukowicz2015}. First we calculated extinction following \citet{nataf2013}:
\begin{equation}
A_{I}=0.7465E(V-I)+1.3700E(J-K)
\end{equation}
where 
\begin{equation}
E(V-I)=(V-I)-(V-I)_{0}.
\end{equation}

The intrinsic colour index $(V-I)_{0}$ is calculated using eq.~\ref{Eq:Colour}, $(V-I)$ is the observed colour, which is the least certain value because $V$-observations contain typically only between several tens and a few hundreds of data points. The reddening $E(J-K)$ comes from \citet{gonzalez2012}. Finally the distance is calculated as

\begin{equation}
d=10^{1-0.2(I-A_{I}-M_{I})}.
\end{equation}

Although the incidence rate of modulated stars slightly differs at different coordinates and distances, we did not detect any significant inhomogeneity, nor clustering of modulated stars (Fig.~\ref{Fig:SpatialDistribution}). Observed amplitudes and periods also do not show any apparent preference for any location or direction. Therefore, we see that BL and single-periodic stars are well mixed and that spatial distribution has no influence on the BL effect.

\section{Summary and conclusions}\label{Sect:Summary}

We performed a detailed, semi-automatic analysis of the frequency spectra of 8282 fundamental mode RRL stars located in the GB aiming to identify the BL effect. To increase the chance to detect the modulation, we used only high-quality observations of the OGLE-IV project and carefully selected stars with suitable data. After prewhitening the spectra of all 8282 stars with ten pulsation harmonics we performed thorough visual inspection of the vicinity of the main pulsation component $f_{0}$ and its harmonic $2f_{0}$ to identify possible side peaks. After identification of BL stars we were able to compare their basic pulsation properties, light curve characteristics and (computed) physical parameters with those of RRL stars without modulation. The main results of our work can be summarized in the following points:
\begin{enumerate}
	\item We identified 3341 stars with modulation, which is almost twice as much as is the number of all catalogued BL stars in all stellar systems so far. In addition, we identified 26 candidate stars and 53 stars showing secular/irregular period changes. 
	\item Half of the stars identified as PC in OGLE-IV data set appeared to be BL stars in combined OGLE-III+OGLE-IV data sets revealing the shortcomings of BL statistics based on dataset with short time span.
	\item The occurrence rate of BL stars is $40.3$\,\%, which is about 10\,\% more than were previous estimates in the GB, and a bit less than are the current estimates for stars in the Galactic field. The true percentage is, however, surely higher than 40.3\,\%. There are a few reasons for that. For example, the incidence rate of identified stars is lower for stars with low number of observations; BL effect with the lowest amplitudes detected in space data could not be detected in available ground-based observations, etc. 
	\item We bring clear and statistically significant evidence that BL stars occur rarely among stars with pulsation periods longer than 0.65\,d. Due to the large analysed sample and good quality data this could hardly be an observational bias (as it could be in some of the publications in literature). Because also other published studies on various studied stellar system suggest similar trend, presence of BL stars only among short-pulsation period RRL stars has very likely general plausibility.
	\item All the mean light curve characteristics of modulated stars related to amplitudes and phases are smaller than for non-modulated stars. The average difference in the total mean amplitudes is linearly period-dependent with the mean value 0.094(13)\,mag. The difference is significantly larger for stars with pulsation periods shorter than 0.425\,d and smaller for stars with periods longer than 0.6\,d.
	\item We derived new interrelations between Fourier coefficients in $I$ passband that allow to calculate the compatibility parameter $D_{m}$. We show that BL stars with $D_{m}>3$ significantly differ from other BL and non-modulated stars in $R_{31}$ vs. $\phi_{31}$ and $\phi_{21}$ vs. $R_{31}$ planes. Regarding this finding, we propose a new approach for identification of BL stars using new empirical polynomial relations (eq. \ref{Eq:Diff1} and \ref{Eq:Diff2}). The portion of the BL stars further than $3\sigma$ below the fitted polynomials is 94\,\%. 
	\item Using empirical relations we computed metallicity, intrinsic colour, and absolute magnitude for stars with $D_{m}<3$. Modulated stars in the GB show the same metallicity, intrinsic $(V-I)_{0}$ colour index and absolute magnitude as stars with nBL light curve suggesting that there is no connection of the occurrence of the BL effect with these parameters in the GB. 
\end{enumerate}	

It is not clear at the moment whether our percentage represents general incidence rate of the BL stars in all stellar systems, or whether it is a unique, intrinsic characteristic of the GB RRL stars.

Our results show that searching for statistical differences between BL and non-BL stars based purely on single-filter photometry cannot help much in solving the BL enigma, even when using very good quality OGLE-IV observations and large number of stars. The (non)presence of the BL effect in a particular RRL star is likely influenced mainly by mass and size of an RRL star because these parameters are directly connected with the pulsation period. Because BL effect is very rare among RRL stars with long-pulsation periods, it could be likely that modulation occurs only in stars that have precisely tuned mass and size, i.e., density. A certain mass-size limit could explain why modulation is common only in short-pulsation period RRL stars and explain the sudden (dis)appearance of the modulation reported in literature for several stars \citep[e.g. RR Gem and V79 M3][]{sodor2007,goranskij2010}. The detail investigation of this idea would need extensive pulsation and evolution modelling complemented with thorough analysis of the observational data, which is certainly out of scope of this general paper. 

Further details about BL stars in GB will be elaborated in the forthcoming papers by \citet[][modulation periods and amplitudes]{skarka2017} and \citet[][the Oosterhoff phenomenon]{prudil2017}.

\section*{Acknowledgements}

We are very grateful to J. Jurcsik and G. Kov\'acs for their very useful comments on the original version of the manuscript, and the anonymous referee for additional suggestions that helped to improve the quality of the paper. We would also like to thank the OGLE team for their great job with the observations of the GB. M.S. acknowledges the support of the postdoctoral fellowship programme of the Hungarian Academy of Sciences at the Konkoly Observatory as a host institution and the financial support of the Hungarian NKFIH Grant K-115709.

\bibliographystyle{mnras}
\bibliography{references}

\begin{thebibliography}{}
\makeatletter
\relax
\def\mn@urlcharsother{\let\do\@makeother \do\$\do\&\do\#\do\^\do\_\do\%\do\~}
\def\mn@doi{\begingroup\mn@urlcharsother \@ifnextchar [ {\mn@doi@}
  {\mn@doi@[]}}
\def\mn@doi@[#1]#2{\def\@tempa{#1}\ifx\@tempa\@empty \href
  {http://dx.doi.org/#2} {doi:#2}\else \href {http://dx.doi.org/#2} {#1}\fi
  \endgroup}
\def\mn@eprint#1#2{\mn@eprint@#1:#2::\@nil}
\def\mn@eprint@arXiv#1{\href {http://arxiv.org/abs/#1} {{\tt arXiv:#1}}}
\def\mn@eprint@dblp#1{\href {http://dblp.uni-trier.de/rec/bibtex/#1.xml}
  {dblp:#1}}
\def\mn@eprint@#1:#2:#3:#4\@nil{\def\@tempa {#1}\def\@tempb {#2}\def\@tempc
  {#3}\ifx \@tempc \@empty \let \@tempc \@tempb \let \@tempb \@tempa \fi \ifx
  \@tempb \@empty \def\@tempb {arXiv}\fi \@ifundefined
  {mn@eprint@\@tempb}{\@tempb:\@tempc}{\expandafter \expandafter \csname
  mn@eprint@\@tempb\endcsname \expandafter{\@tempc}}}

\bibitem[\protect\citeauthoryear{{Alcock} et~al.,}{{Alcock}
  et~al.}{2003}]{alcock2003}
{Alcock} C.,  et~al., 2003, \mn@doi [\apj] {10.1086/378689}, \href
  {http://adsabs.harvard.edu/abs/2003ApJ...598..597A} {598, 597}

\bibitem[\protect\citeauthoryear{{Arellano Ferro}, {Bramich}, {Figuera Jaimes},
  {Giridhar}  \& {Kuppuswamy}}{{Arellano Ferro} et~al.}{2012}]{arellano2012}
{Arellano Ferro} A.,  {Bramich} D.~M.,  {Figuera Jaimes} R.,  {Giridhar} S.,
  {Kuppuswamy} K.,  2012, \mn@doi [\mnras] {10.1111/j.1365-2966.2011.20119.x},
  \href {http://adsabs.harvard.edu/abs/2012MNRAS.420.1333A} {420, 1333}

\bibitem[\protect\citeauthoryear{{Benk{\H o}}, {Plachy}, {Szab{\'o}},
  {Moln{\'a}r}  \& {Koll{\'a}th}}{{Benk{\H o}} et~al.}{2014}]{benko2014}
{Benk{\H o}} J.~M.,  {Plachy} E.,  {Szab{\'o}} R.,  {Moln{\'a}r} L.,
  {Koll{\'a}th} Z.,  2014, \mn@doi [\apjs] {10.1088/0067-0049/213/2/31}, \href
  {http://adsabs.harvard.edu/abs/2014ApJS..213...31B} {213, 31}

\bibitem[\protect\citeauthoryear{{Benk{\H o}}, {Szab{\'o}}, {Derekas}  \&
  {S{\'o}dor}}{{Benk{\H o}} et~al.}{2016}]{benko2016}
{Benk{\H o}} J.~M.,  {Szab{\'o}} R.,  {Derekas} A.,   {S{\'o}dor} {\'A}.,
  2016, \mn@doi [\mnras] {10.1093/mnras/stw2136}, \href
  {http://adsabs.harvard.edu/abs/2016MNRAS.463.1769B} {463, 1769}

\bibitem[\protect\citeauthoryear{{Bla{\v z}ko}}{{Bla{\v
  z}ko}}{1907}]{blazhko1907}
{Bla{\v z}ko} S.,  1907, \mn@doi [Astronomische Nachrichten]
  {10.1002/asna.19071752002}, \href
  {http://adsabs.harvard.edu/abs/1907AN....175..325B} {175, 325}

\bibitem[\protect\citeauthoryear{{Buchler} \& {Koll{\'a}th}}{{Buchler} \&
  {Koll{\'a}th}}{2011}]{buchler2011}
{Buchler} J.~R.,  {Koll{\'a}th} Z.,  2011, \mn@doi [\apj]
  {10.1088/0004-637X/731/1/24}, \href
  {http://adsabs.harvard.edu/abs/2011ApJ...731...24B} {731, 24}

\bibitem[\protect\citeauthoryear{{Cacciari}, {Corwin}  \& {Carney}}{{Cacciari}
  et~al.}{2005}]{cacciari2005}
{Cacciari} C.,  {Corwin} T.~M.,   {Carney} B.~W.,  2005, \mn@doi [\aj]
  {10.1086/426325}, \href {http://adsabs.harvard.edu/abs/2005AJ....129..267C}
  {129, 267}

\bibitem[\protect\citeauthoryear{{Catelan} \& {Cort{\'e}s}}{{Catelan} \&
  {Cort{\'e}s}}{2008}]{catelan2008}
{Catelan} M.,  {Cort{\'e}s} C.,  2008, \mn@doi [\apjl] {10.1086/587515}, \href
  {http://adsabs.harvard.edu/abs/2008ApJ...676L.135C} {676, L135}

\bibitem[\protect\citeauthoryear{{Catelan}, {Pritzl}  \& {Smith}}{{Catelan}
  et~al.}{2004}]{catelan2004}
{Catelan} M.,  {Pritzl} B.~J.,   {Smith} H.~A.,  2004, \mn@doi [\apjs]
  {10.1086/422916}, \href {http://adsabs.harvard.edu/abs/2004ApJS..154..633C}
  {154, 633}

\bibitem[\protect\citeauthoryear{{Collinge}, {Sumi}  \& {Fabrycky}}{{Collinge}
  et~al.}{2006}]{collinge2006}
{Collinge} M.~J.,  {Sumi} T.,   {Fabrycky} D.,  2006, \mn@doi [\apj]
  {10.1086/507407}, \href {http://adsabs.harvard.edu/abs/2006ApJ...651..197C}
  {651, 197}

\bibitem[\protect\citeauthoryear{{Gonzalez}, {Rejkuba}, {Zoccali}, {Valenti},
  {Minniti}, {Schultheis}, {Tobar}  \& {Chen}}{{Gonzalez}
  et~al.}{2012}]{gonzalez2012}
{Gonzalez} O.~A.,  {Rejkuba} M.,  {Zoccali} M.,  {Valenti} E.,  {Minniti} D.,
  {Schultheis} M.,  {Tobar} R.,   {Chen} B.,  2012, \mn@doi [\aap]
  {10.1051/0004-6361/201219222}, \href
  {http://adsabs.harvard.edu/abs/2012A%26A...543A..13G} {543, A13}

\bibitem[\protect\citeauthoryear{{Goranskij}, {Clement}  \&
  {Thompson}}{{Goranskij} et~al.}{2010}]{goranskij2010}
{Goranskij} V.,  {Clement} C.~M.,   {Thompson} M.,  2010, in {Sterken} C.,
  {Samus} N.,   {Szabados} L.,  eds, Variable Stars, the Galactic halo and
  Galaxy Formation.

\bibitem[\protect\citeauthoryear{{Hajdu}, {Catelan}, {Jurcsik},
  {D{\'e}k{\'a}ny}, {Drake}  \& {Marquette}}{{Hajdu} et~al.}{2015}]{hajdu2015}
{Hajdu} G.,  {Catelan} M.,  {Jurcsik} J.,  {D{\'e}k{\'a}ny} I.,  {Drake} A.~J.,
    {Marquette} J.-B.,  2015, \mn@doi [\mnras] {10.1093/mnrasl/slv024}, \href
  {http://adsabs.harvard.edu/abs/2015MNRAS.449L.113H} {449, L113}

\bibitem[\protect\citeauthoryear{{Howell} et~al.,}{{Howell}
  et~al.}{2014}]{howell2014}
{Howell} S.~B.,  et~al., 2014, \mn@doi [\pasp] {10.1086/676406}, \href
  {http://adsabs.harvard.edu/abs/2014PASP..126..398H} {126, 398}

\bibitem[\protect\citeauthoryear{{Jurcsik} \& {Kovacs}}{{Jurcsik} \&
  {Kovacs}}{1996}]{jurcsik1996}
{Jurcsik} J.,  {Kovacs} G.,  1996, \aap, \href
  {http://adsabs.harvard.edu/abs/1996A%26A...312..111J} {312, 111}

\bibitem[\protect\citeauthoryear{{Jurcsik} \& {Smitola}}{{Jurcsik} \&
  {Smitola}}{2016}]{jurcsik2016}
{Jurcsik} J.,  {Smitola} P.,  2016, Commmunications of the Konkoly Observatory
  Hungary, \href {http://adsabs.harvard.edu/abs/2016CoKon.105..167J} {105, 167}

\bibitem[\protect\citeauthoryear{{Jurcsik}, {Sodor}  \& {Varadi}}{{Jurcsik}
  et~al.}{2005}]{jurcsik2005}
{Jurcsik} J.,  {Sodor} A.,   {Varadi} M.,  2005, Information Bulletin on
  Variable Stars, \href {http://adsabs.harvard.edu/abs/2005IBVS.5666....1J}
  {5666}

\bibitem[\protect\citeauthoryear{{Jurcsik} et~al.,}{{Jurcsik}
  et~al.}{2009}]{jurcsik2009}
{Jurcsik} J.,  et~al., 2009, \mn@doi [\mnras]
  {10.1111/j.1365-2966.2009.15515.x}, \href
  {http://adsabs.harvard.edu/abs/2009MNRAS.400.1006J} {400, 1006}

\bibitem[\protect\citeauthoryear{{Jurcsik}, {Szeidl}, {Clement}, {Hurta}  \&
  {Lovas}}{{Jurcsik} et~al.}{2011}]{jurcsik2011}
{Jurcsik} J.,  {Szeidl} B.,  {Clement} C.,  {Hurta} Z.,   {Lovas} M.,  2011,
  \mn@doi [\mnras] {10.1111/j.1365-2966.2010.17817.x}, \href
  {http://adsabs.harvard.edu/abs/2011MNRAS.411.1763J} {411, 1763}

\bibitem[\protect\citeauthoryear{{Jurcsik} et~al.,}{{Jurcsik}
  et~al.}{2012}]{jurcsik2012}
{Jurcsik} J.,  et~al., 2012, \mn@doi [\mnras]
  {10.1111/j.1365-2966.2011.19868.x}, \href
  {http://adsabs.harvard.edu/abs/2012MNRAS.419.2173J} {419, 2173}

\bibitem[\protect\citeauthoryear{{Koll{\'a}th}, {Moln{\'a}r}  \&
  {Szab{\'o}}}{{Koll{\'a}th} et~al.}{2011}]{kollath2011}
{Koll{\'a}th} Z.,  {Moln{\'a}r} L.,   {Szab{\'o}} R.,  2011, \mn@doi [\mnras]
  {10.1111/j.1365-2966.2011.18451.x}, \href
  {http://adsabs.harvard.edu/abs/2011MNRAS.414.1111K} {414, 1111}

\bibitem[\protect\citeauthoryear{{Kov\'{a}cs}}{{Kov\'{a}cs}}{2016}]{kovacs2016}
{Kov\'{a}cs} G.,  2016, Communications of the Konkoly Observatory Hungary,
  \href {http://adsabs.harvard.edu/abs/2016CoKon.105...61K} {105, 61}

\bibitem[\protect\citeauthoryear{{Lenz} \& {Breger}}{{Lenz} \&
  {Breger}}{2004}]{lenz2004}
{Lenz} P.,  {Breger} M.,  2004, in {Zverko} J.,  {Ziznovsky} J.,  {Adelman}
  S.~J.,   {Weiss} W.~W.,  eds,  IAU Symposium Vol. 224, The A-Star Puzzle. pp
  786--790, \mn@doi{10.1017/S1743921305009750}

\bibitem[\protect\citeauthoryear{{Mizerski}}{{Mizerski}}{2003}]{mizerski2003}
{Mizerski} T.,  2003, \actaa, \href
  {http://adsabs.harvard.edu/abs/2003AcA....53..307M} {53, 307}

\bibitem[\protect\citeauthoryear{{Moskalik} \& {Poretti}}{{Moskalik} \&
  {Poretti}}{2003}]{moskalik2003}
{Moskalik} P.,  {Poretti} E.,  2003, \mn@doi [\aap]
  {10.1051/0004-6361:20021595}, \href
  {http://adsabs.harvard.edu/abs/2003A%26A...398..213M} {398, 213}

\bibitem[\protect\citeauthoryear{{Nagy} \& {Kov{\'a}cs}}{{Nagy} \&
  {Kov{\'a}cs}}{2006}]{nagy2006}
{Nagy} A.,  {Kov{\'a}cs} G.,  2006, \mn@doi [\aap]
  {10.1051/0004-6361:20054538}, \href
  {http://adsabs.harvard.edu/abs/2006A%26A...454..257N} {454, 257}

\bibitem[\protect\citeauthoryear{{Nataf} et~al.,}{{Nataf}
  et~al.}{2013}]{nataf2013}
{Nataf} D.~M.,  et~al., 2013, \mn@doi [\apj] {10.1088/0004-637X/769/2/88},
  \href {http://adsabs.harvard.edu/abs/2013ApJ...769...88N} {769, 88}

\bibitem[\protect\citeauthoryear{{Nemec}, {Cohen}, {Ripepi}, {Derekas},
  {Moskalik}, {Sesar}, {Chadid}  \& {Bruntt}}{{Nemec} et~al.}{2013}]{nemec2013}
{Nemec} J.~M.,  {Cohen} J.~G.,  {Ripepi} V.,  {Derekas} A.,  {Moskalik} P.,
  {Sesar} B.,  {Chadid} M.,   {Bruntt} H.,  2013, \mn@doi [\apj]
  {10.1088/0004-637X/773/2/181}, \href
  {http://adsabs.harvard.edu/abs/2013ApJ...773..181N} {773, 181}

\bibitem[\protect\citeauthoryear{{Pietrukowicz} et~al.,}{{Pietrukowicz}
  et~al.}{2015}]{pietrukowicz2015}
{Pietrukowicz} P.,  et~al., 2015, \mn@doi [\apj] {10.1088/0004-637X/811/2/113},
  \href {http://adsabs.harvard.edu/abs/2015ApJ...811..113P} {811, 113}

\bibitem[\protect\citeauthoryear{{Prudil}, {Smolec}, {Skarka}  \&
  {Netzel}}{{Prudil} et~al.}{2016}]{prudil2016}
{Prudil} Z.,  {Smolec} R.,  {Skarka} M.,   {Netzel} H.,  2016, MNRAS, submitted

\bibitem[\protect\citeauthoryear{{Prudil}, {Skarka}  \& {et al.,}}{{Prudil}
  et~al.}{2017}]{prudil2017}
{Prudil} Z.,  {Skarka} M.,   {et al.,} 2017, in prep.

\bibitem[\protect\citeauthoryear{{Simon} \& {Lee}}{{Simon} \&
  {Lee}}{1981}]{simon1981}
{Simon} N.~R.,  {Lee} A.~S.,  1981, \mn@doi [\apj] {10.1086/159153}, \href
  {http://adsabs.harvard.edu/abs/1981ApJ...248..291S} {248, 291}

\bibitem[\protect\citeauthoryear{{Skarka}}{{Skarka}}{2014a}]{skarka2014b}
{Skarka} M.,  2014a, \mn@doi [\mnras] {10.1093/mnras/stu1815}, \href
  {http://adsabs.harvard.edu/abs/2014MNRAS.445.1584S} {445, 1584}

\bibitem[\protect\citeauthoryear{{Skarka}}{{Skarka}}{2014b}]{skarka2014a}
{Skarka} M.,  2014b, \mn@doi [\aap] {10.1051/0004-6361/201322491}, \href
  {http://adsabs.harvard.edu/abs/2014A%26A...562A..90S} {562, A90}

\bibitem[\protect\citeauthoryear{{Skarka}, {Li{\v s}ka}, {Auer}, {Prudil},
  {Jur{\'a}{\v n}ov{\'a}}  \& {S{\'o}dor}}{{Skarka} et~al.}{2016}]{skarka2016b}
{Skarka} M.,  {Li{\v s}ka} J.,  {Auer} R.~F.,  {Prudil} Z.,  {Jur{\'a}{\v
  n}ov{\'a}} A.,   {S{\'o}dor} {\'A}.,  2016, \mn@doi [\aap]
  {10.1051/0004-6361/201628855}, \href
  {http://adsabs.harvard.edu/abs/2016A%26A...592A.144S} {592, A144}

\bibitem[\protect\citeauthoryear{{Skarka}, {Prudil}  \& {et al.,}}{{Skarka}
  et~al.}{2017}]{skarka2017}
{Skarka} M.,  {Prudil} Z.,   {et al.,} 2017, in prep.

\bibitem[\protect\citeauthoryear{{Smolec}}{{Smolec}}{2005}]{smolec2005}
{Smolec} R.,  2005, \actaa, \href
  {http://adsabs.harvard.edu/abs/2005AcA....55...59S} {55, 59}

\bibitem[\protect\citeauthoryear{{Smolec}, {Prudil}, {Skarka}  \&
  {Bakowska}}{{Smolec} et~al.}{2016}]{smolec2016}
{Smolec} R.,  {Prudil} Z.,  {Skarka} M.,   {Bakowska} K.,  2016, \mn@doi
  [\mnras] {10.1093/mnras/stw1519}, \href
  {http://adsabs.harvard.edu/abs/2016MNRAS.461.2934S} {461, 2934}

\bibitem[\protect\citeauthoryear{{S{\'o}dor}, {Szeidl}  \&
  {Jurcsik}}{{S{\'o}dor} et~al.}{2007}]{sodor2007}
{S{\'o}dor} {\'A}.,  {Szeidl} B.,   {Jurcsik} J.,  2007, \mn@doi [\aap]
  {10.1051/0004-6361:20066886}, \href
  {http://adsabs.harvard.edu/abs/2007A%26A...469.1033S} {469, 1033}

\bibitem[\protect\citeauthoryear{{S{\'o}dor} et~al.,}{{S{\'o}dor}
  et~al.}{2012}]{sodor2012}
{S{\'o}dor} {\'A}.,  et~al., 2012, in {Shibahashi} H.,  {Takata} M.,
  {Lynas-Gray} A.~E.,  eds,  Astronomical Society of the Pacific Conference
  Series Vol. 462, Progress in Solar/Stellar Physics with Helio- and
  Asteroseismology. p.~228

\bibitem[\protect\citeauthoryear{{Soszy{\'n}ski} et~al.,}{{Soszy{\'n}ski}
  et~al.}{2011}]{soszynski2011}
{Soszy{\'n}ski} I.,  et~al., 2011, \actaa, \href
  {http://adsabs.harvard.edu/abs/2011AcA....61....1S} {61, 1}

\bibitem[\protect\citeauthoryear{{Soszy{\'n}ski} et~al.,}{{Soszy{\'n}ski}
  et~al.}{2014}]{soszynski2014}
{Soszy{\'n}ski} I.,  et~al., 2014, \actaa, \href
  {http://adsabs.harvard.edu/abs/2014AcA....64..177S} {64, 177}

\bibitem[\protect\citeauthoryear{{Soszy{\'n}ski} et~al.,}{{Soszy{\'n}ski}
  et~al.}{2016}]{soszynski2016}
{Soszy{\'n}ski} I.,  et~al., 2016, \actaa, \href
  {http://adsabs.harvard.edu/abs/2016AcA....66..131S} {66, 131}

\bibitem[\protect\citeauthoryear{{Szab{\'o}} et~al.,}{{Szab{\'o}}
  et~al.}{2010}]{szabo2010}
{Szab{\'o}} R.,  et~al., 2010, \mn@doi [\mnras]
  {10.1111/j.1365-2966.2010.17386.x}, \href
  {http://adsabs.harvard.edu/abs/2010MNRAS.409.1244S} {409, 1244}

\bibitem[\protect\citeauthoryear{{Szab{\'o}} et~al.,}{{Szab{\'o}}
  et~al.}{2014}]{szabo2014}
{Szab{\'o}} R.,  et~al., 2014, \mn@doi [\aap] {10.1051/0004-6361/201424522},
  \href {http://adsabs.harvard.edu/abs/2014A%26A...570A.100S} {570, A100}

\bibitem[\protect\citeauthoryear{{Szczygie{\l}} \& {Fabrycky}}{{Szczygie{\l}}
  \& {Fabrycky}}{2007}]{szczygiel2007}
{Szczygie{\l}} D.~M.,  {Fabrycky} D.~C.,  2007, \mn@doi [\mnras]
  {10.1111/j.1365-2966.2007.11678.x}, \href
  {http://adsabs.harvard.edu/abs/2007MNRAS.377.1263S} {377, 1263}

\bibitem[\protect\citeauthoryear{{Szeidl}}{{Szeidl}}{1988}]{szeidl1988}
{Szeidl} B.,  1988, in {Kovacs} G.,  {Szabados} L.,   {Szeidl} B.,  eds,
  Multimode Stellar Pulsations. p.~45

\bibitem[\protect\citeauthoryear{{Szeidl} \& {Jurcsik}}{{Szeidl} \&
  {Jurcsik}}{2009}]{szeidl2009}
{Szeidl} B.,  {Jurcsik} J.,  2009, Communications in Asteroseismology, \href
  {http://adsabs.harvard.edu/abs/2009CoAst.160...17S} {160, 17}

\bibitem[\protect\citeauthoryear{{Szeidl}, {Hurta}, {Jurcsik}, {Clement}  \&
  {Lovas}}{{Szeidl} et~al.}{2011}]{szeidl2011}
{Szeidl} B.,  {Hurta} Z.,  {Jurcsik} J.,  {Clement} C.,   {Lovas} M.,  2011,
  \mn@doi [\mnras] {10.1111/j.1365-2966.2010.17815.x}, \href
  {http://adsabs.harvard.edu/abs/2011MNRAS.411.1744S} {411, 1744}

\bibitem[\protect\citeauthoryear{{Udalski}}{{Udalski}}{2003}]{udalski2003}
{Udalski} A.,  2003, \actaa, \href
  {http://adsabs.harvard.edu/abs/2003AcA....53..291U} {53, 291}

\bibitem[\protect\citeauthoryear{{Udalski}, {Szymanski}, {Kaluzny}, {Kubiak}
  \& {Mateo}}{{Udalski} et~al.}{1992}]{udalski1992}
{Udalski} A.,  {Szymanski} M.,  {Kaluzny} J.,  {Kubiak} M.,   {Mateo} M.,
  1992, \actaa, \href {http://adsabs.harvard.edu/abs/1992AcA....42..253U} {42,
  253}

\bibitem[\protect\citeauthoryear{{Udalski}, {Kubiak}  \& {Szymanski}}{{Udalski}
  et~al.}{1997}]{udalski1997}
{Udalski} A.,  {Kubiak} M.,   {Szymanski} M.,  1997, \actaa, \href
  {http://adsabs.harvard.edu/abs/1997AcA....47..319U} {47, 319}

\bibitem[\protect\citeauthoryear{{Udalski}, {Szyma{\'n}ski}  \&
  {Szyma{\'n}ski}}{{Udalski} et~al.}{2015}]{udalski2015}
{Udalski} A.,  {Szyma{\'n}ski} M.~K.,   {Szyma{\'n}ski} G.,  2015, \actaa,
  \href {http://adsabs.harvard.edu/abs/2015AcA....65....1U} {65, 1}

\makeatother
\end{thebibliography}

\bsp	
\label{lastpage}
\end{document}